 \documentclass[12pt,draftcls,onecolumn,twoside]{IEEEtran}
 
\usepackage{amsmath}

\usepackage{bm}

\DeclareMathAlphabet{\mathsfbr}{OT1}{cmss}{m}{n}
\SetMathAlphabet{\mathsfbr}{bold}{OT1}{cmss}{bx}{n}
\DeclareRobustCommand{\msf}[1]{%
  \ifcat\noexpand#1\relax\msfgreek{#1}\else\mathsfbr{#1}\fi
}

\makeatletter
\newcommand{\msfgreek}[1]{\csname s\expandafter\@gobble\string#1\endcsname}
\makeatother

\DeclareFontEncoding{LGR}{}{} 
\DeclareSymbolFont{sfgreek}{LGR}{cmss}{m}{n}
\SetSymbolFont{sfgreek}{bold}{LGR}{cmss}{bx}{n}
\DeclareMathSymbol{\salpha}{\mathord}{sfgreek}{`a}
\DeclareMathSymbol{\sbeta}{\mathord}{sfgreek}{`b}
\DeclareMathSymbol{\sgamma}{\mathord}{sfgreek}{`g}
\DeclareMathSymbol{\sdelta}{\mathord}{sfgreek}{`d}
\DeclareMathSymbol{\sepsilon}{\mathord}{sfgreek}{`e}
\DeclareMathSymbol{\szeta}{\mathord}{sfgreek}{`z}
\DeclareMathSymbol{\seta}{\mathord}{sfgreek}{`h}
\DeclareMathSymbol{\stheta}{\mathord}{sfgreek}{`j}
\DeclareMathSymbol{\siota}{\mathord}{sfgreek}{`i}
\DeclareMathSymbol{\skappa}{\mathord}{sfgreek}{`k}
\DeclareMathSymbol{\slambda}{\mathord}{sfgreek}{`l}
\DeclareMathSymbol{\smu}{\mathord}{sfgreek}{`m}
\DeclareMathSymbol{\snu}{\mathord}{sfgreek}{`n}
\DeclareMathSymbol{\sxi}{\mathord}{sfgreek}{`x}
\DeclareMathSymbol{\somicron}{\mathord}{sfgreek}{`o}
\DeclareMathSymbol{\spi}{\mathord}{sfgreek}{`p}
\DeclareMathSymbol{\srho}{\mathord}{sfgreek}{`r}
\DeclareMathSymbol{\ssigma}{\mathord}{sfgreek}{`s}
\DeclareMathSymbol{\stau}{\mathord}{sfgreek}{`t}
\DeclareMathSymbol{\supsilon}{\mathord}{sfgreek}{`u}
\DeclareMathSymbol{\sphi}{\mathord}{sfgreek}{`f}
\DeclareMathSymbol{\schi}{\mathord}{sfgreek}{`q}
\DeclareMathSymbol{\spsi}{\mathord}{sfgreek}{`y}
\DeclareMathSymbol{\somega}{\mathord}{sfgreek}{`w}

\DeclareMathSymbol{\svarsigma}{\mathord}{sfgreek}{`c}

\DeclareMathSymbol{\sGamma}{\mathalpha}{sfgreek}{`G}
\DeclareMathSymbol{\sDelta}{\mathalpha}{sfgreek}{`D}
\DeclareMathSymbol{\sTheta}{\mathalpha}{sfgreek}{`J}
\DeclareMathSymbol{\sLambda}{\mathalpha}{sfgreek}{`L}
\DeclareMathSymbol{\sXi}{\mathalpha}{sfgreek}{`X}
\DeclareMathSymbol{\sPi}{\mathalpha}{sfgreek}{`P}
\DeclareMathSymbol{\sSigma}{\mathalpha}{sfgreek}{`S}
\DeclareMathSymbol{\sUpsilon}{\mathalpha}{sfgreek}{`U}
\DeclareMathSymbol{\sPhi}{\mathalpha}{sfgreek}{`F}
\DeclareMathSymbol{\sPsi}{\mathalpha}{sfgreek}{`Y}
\DeclareMathSymbol{\sOmega}{\mathalpha}{sfgreek}{`W}

\DeclareRobustCommand{\mcal}[1]{%
  \ifcat\noexpand#1\relax\mathnormal{#1}\else\cal{#1}\fi
}
\DeclareRobustCommand{\BM}[1]{%
  \ifcat\noexpand#1\relax\bm{\boldUppercaseItalicGreek{#1}}\else\bm{#1}\fi
}
\makeatletter
\newcommand{\boldUppercaseItalicGreek}[1]{\csname var\expandafter\@gobble\string#1\endcsname}
\makeatother

\newcommand{\V}[1]{\bm{#1}}
\newcommand{\M}[1]{\BM{#1}}
\newcommand{\Set}[1]{\mathcal{#1}}

\usepackage{acronym,mathrsfs,dsfont,amssymb}  
\usepackage{color,cite}  
\usepackage{graphicx}%
\usepackage{physics}
\usepackage{xtab,booktabs}
\usepackage[tableposition=top]{caption}
\usepackage{longtable}
\usepackage{colortbl,pgfplotstable}
\usepackage{caption}
\usepackage{subcaption}

\newcommand{\bd}{\begin{description}}
\newcommand{\ed}{\end{description}}
\newcommand{\be}{\begin{enumerate}}
\newcommand{\ee}{\end{enumerate}}
\newcommand{\bi}{\begin{itemize}}
\newcommand{\ei}{\end{itemize}}
\newcommand{\bl}{\begin{list}}
\newcommand{\el}{\end{list}}
\newcommand{\bt}{\begin{tabbing}}
\newcommand{\et}{\end{tabbing}}

\definecolor{BLUE}{rgb}{0,0,1}

\newtheorem{definition}{Definition}
\newtheorem{theorem}{Theorem}
\newtheorem{lemma}{Lemma}

\newcommand\Cb[1] {{{#1}}}

\acrodef{epr}[EPR]{Einstein-Podolsky-Rosen}
\acrodef{fifo}[FIFO]{First-In-First-Out}
\acrodef{oqf}[OQF]{Oldest-Qubit-First}
\acrodef{yqf}[YQF]{youngest-Qubit-First}

\newcommand\BSN{W} 
\newcommand\Ent{E} 
\newcommand\A{A} 
\newcommand\WA{\widetilde{A}} 
\newcommand\UU{U} 
\newcommand\WU{\widetilde{U}} 
\newcommand\F{F} 
\newcommand\RF{\mu} 
\newcommand\CC{C}

\begin{document}

\title{Entanglement Swapping in Quantum Switches: Protocol Design and Stability Analysis}





\author{
	\vspace{0.2cm}
    Wenhan~Dai,
    Anthony~Rinaldi, and
    Don~Towsley  
    \thanks{
	The material in this paper was presented, in part,
		at the IEEE International Conference on Quantum Computing and Engineering, Broomfield, CO, USA, September 2022.}
\thanks{W.~Dai and D.~Towsley are with the College of Information and Computer Sciences, University of Massachusetts, Amherst (e-mail:  whdai@cs.umass.edu and towsley@cs.umass.edu).}
\thanks{A.~Rinaldi is with University of Massachusetts, Amherst (e-mail:  aarinaldi@umass.edu).}
    }

%
%

\maketitle 

\setcounter{page}{1}

\begin{abstract}
Quantum switches are critical components in quantum networks, which distribute entangled pairs among end nodes by entanglement swapping. In this work, we design protocols that schedule entanglement swapping operations in a quantum switch. Entanglement requests randomly arrive at the switch, and the goal of an entanglement swapping protocol is to stabilize the quantum switch so that the number of unfinished entanglement requests is bounded with a high probability. We determine the capacity region for the rates of entanglement requests that the quantum switch can stably support. We also develop protocols that not only stabilize the switch,  but also achieve zero average latency. Among these protocols, the on-demand protocols are not only computationally efficient, but also achieve high fidelity and low latency demonstrated by results obtained using a quantum network discrete event simulator.

\end{abstract}



\begin{IEEEkeywords}
quantum switch, entanglement distribution, quantum networking
\end{IEEEkeywords}

\maketitle

%


\acresetall		

\section{Introduction}\label{sec:intro}
 
 Quantum networks {will} play a critical role in enabling numerous  quantum applications such as quantum key distribution \cite{ShoPre:00,LoCha:99,GotLoLutPre:04}, teleportation \cite{BenBraCreJozPerWoo:93,MaHerSchWanKroNayWitMecKofAni:12, PfaHenBerDamBloTamTigSchMarTwiHan:14}, and quantum sensing \cite{DArLopPar:01, HuaMacMac:16, DemMac:14}. One of the major tasks of quantum networks is distributing quantum entanglement among geographically separated nodes. Such a task usually involves generating \ac{epr} pairs through quantum channels and then performing entanglement swapping among the generated \ac{epr} pairs. For example, consider a star-shape network consisting of a center node and a collection of end nodes. Entanglement swapping is performed at the center node to establish entanglement among end nodes. The center node serves as a quantum switch, a critical building block in quantum networks. See Figure 1 for details.

\begin{figure}[t]
\center	
\includegraphics[width=0.65\linewidth, draft=false]{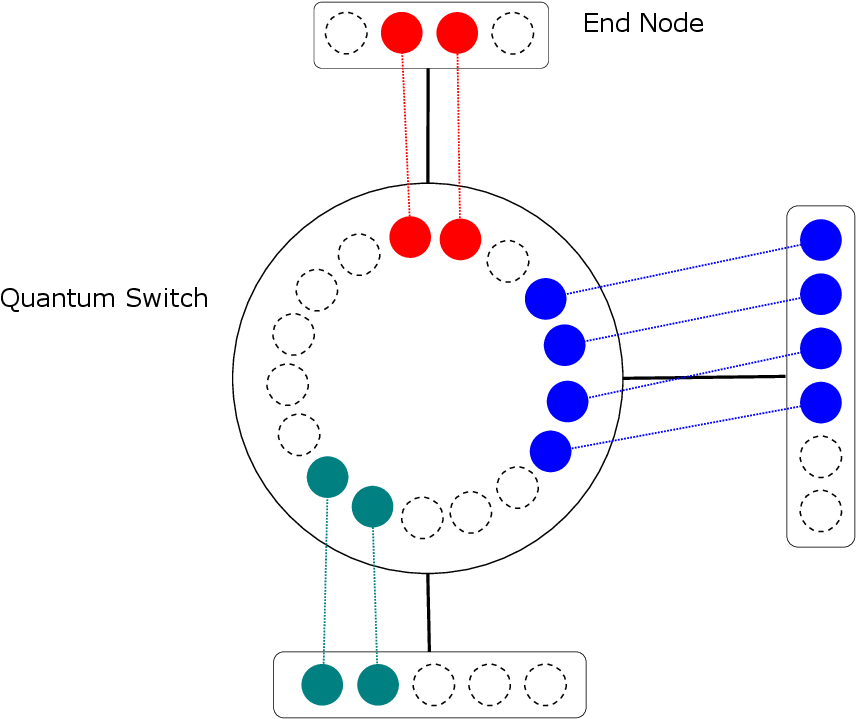}
\caption{Illustration for a quantum switch. The big circle represents the switch and the rectangles represent end nodes. Solid colorful dots represent entangled qubits, and empty dots represent empty memory slots in the quantum switch or end nodes. Different colors correspond to different end nodes. The lines connecting the switch and end nodes represent quantum channels. A colorful dashed line implies that the two dots connected by the dashed line consist of an \ac{epr} pair.}
\label{fig:quantum_switch}
\end{figure}

 
 A key problem in the implementation of a quantum switch is decision-making about which \ac{epr} pairs to perform entanglement swapping operations on. The prioritization of entanglement swapping affects the performance of the switch, such as the fidelity of the distributed entanglement, the latency of the entanglement requests, and the throughput of the switch. Existing studies on entanglement swapping generally focus on maximizing entanglement generation rate, and the quantum network establishes entanglement whenever possible. Relevant work is summarized in Section \ref{sec:relatedwork}. In this manuscript, we consider a more generic and practical scenario where entanglement requests randomly arrive at the switch, and the switch aims to address these requests. Instead of maximizing entanglement generation rate, we tackle seemingly more difficult problems that involve the concept of stability. Roughly speaking, a quantum switch being stable implies that the number of unaddressed entanglement requests is not very large with a high probability. Note that a protocol that maximizes the entanglement generation rate may not stabilize the switch since the number of unaddressed entanglement requests may grow to infinity sub-linearly with respect to time. The goal of this manuscript is to solve the following two problems: What is the capacity region of the switch for the entanglement generation rates that the quantum switch can stably support? How do we design entanglement swapping protocols that achieve high performance for any workload vector within this capacity region?

 
 Note that in classical data networks, stability is widely used to investigate protocols for data routing and resource allocation \cite{TasEph:92, NeeModRoh:05,Tass:93}. These studies cannot be directly applied to quantum networks because entanglement swapping involve two interfaces, whereas data transmission in classical data networks normally involves one link. Moreover, entanglement generated through quantum channels can be stored to satisfy future entanglement requests whereas in classical networks, data have to arrive first and then get transmitted through channels. Though these fundamental differences precludes our directly applying existing results in classical networks, the mathematical tools, such as Lyapunov drift analysis, used to obtain them are useful for the analysis of the entanglement swapping protocols in a quantum switch. With different Lyapunov functions tailored for the quantum switch, we can develop several protocols that not only stabilize the switch, but also achieve zero average latency. The key contributions of this manuscript are as follows:
 \begin{itemize}
 \item We determine the capacity region for the entanglement rates. In particular, we show that if the entanglement rates are outside of this region, no entanglement swapping protocol can stabilize the switch.
 \item For any entanglement rates being an interior point of the capacity region, we develop stationary protocols that can stabilize the switch.
 \item We develop a family of protocols, referred to as on-demand protocols, that can stabilize the switch. These protocols are computationally efficient and do not require statistical knowledge of the entanglement requests and the quantum channels.
 \item We further show that the stationary protocols and on-demand protocols with virtual requests can achieve zero average latency. This means that almost all the requests can be served immediately when they arrive at the switch.
 \item We evaluate the proposed protocols with a quantum network discrete event simulator NetSquid. We compare the protocols according to the fidelity and latency. Moreover, we show the performance of the quantum switch with respect to different factors, such as the memory size, decoherence time, and entanglement swapping success probability. Such performance providers insight into the implementation of quantum switches in noisy intermediate-scale quantum (NISQ) era.
 \end{itemize}

\section{Background}
In this section, we provide some background information that will be used in this manuscript. An \ac{epr} pair is a  quantum state consisting of two qubits:
\begin{align*}
    \ket{\Psi_{AB}} = \frac{1}{\sqrt{2}}(\ket{0_{A}}\ket{0_{B}}+\ket{1_{A}}\ket{1_{B}})
\end{align*}
where $\ket{0}$ and $\ket{1}$ are qubits represented by two-dimensional vectors, and the subscripts $A$ and $B$ represent two physics systems. The \ac{epr} pairs play a critical role in applications such as superdense coding and quantum teleportation.

To generate an \ac{epr} pair between two parties $A$ and $B$ that are apart, one can locally prepare the \ac{epr} locally at one party (for example, a pair of photons that encode qubits with polarization) and then send one qubit to the other party through a quantum channel. Alternatively, one can use entanglement swapping at an intermediate party $C$ to achieve so. Specifically, suppose $A$ and $C$ share an \ac{epr} pair $\ket{\Psi_{AC}}$ and $B$ and $C$ share an \ac{epr} pair $\ket{\Psi_{BC}}$. Then $C$ can make Bell state measurements on its two qubits (one entangled with $A$ and the other entangled with $B$) and send the results to $A$, and $A$ performs one of the Pauli operators on its qubit based on the measurement results from $C$. This operation is known as entanglement swapping \cite{ZukZeiHorEke:93} and its outcome is the \ac{epr} pair $\ket{\Psi_{AB}}$ at the cost of two \ac{epr} pairs $\ket{\Psi_{AC}}$ and  $\ket{\Psi_{BC}}$. In the setup of this manuscript, the quantum switch generates \ac{epr} pairs with end nodes through quantum channels, and perform entanglement swapping to generate \ac{epr} pairs between end nodes.

Generating \ac{epr} pairs between the quantum switch and end nodes requires qubit transmission through quantum channels. One of the widely used medium for qubit transmission is optical fiber, and correspondingly, the quantum information in qubits are carried by photons. For a single photon that goes through optical fiber, with probability $p$ this photon successfully reaches the receiver, and with probability $1-p$ it is lost. The probability $p$ depends on the distance between the transmitter and the receiver and the fiber attenuation coefficient. Using this channel model, we assume that the entanglement generation between the switch and an end node succeeds with a known probability. Note that we consider heralded entanglement generation, i.e., the results of entanglement generation, either success or failure, are known to the switch.

Entanglement swapping requires making Bell state measurements on two qubits at the quantum switch. This can be done by performing a CNOT operation and making measurements with standard computation basis. In practice, the CNOT operation are not always successful if it is implemented by linear optics \cite{EweLoo:14, Gri:11, KokMunNemRalDowMil:07} or photon-spin interaction \cite{DuaKim:04}. Therefore, we model the Bell state measurement as a probabilistic operator: with probability $q$, it succeeds and the  \ac{epr} pair between the corresponding two end nodes is generated; with probability $1-q$, no \ac{epr} pair is generated between the two end nodes although the two \ac{epr} pairs between the quantum switch and the two end nodes are consumed.

\section{Related Work}\label{sec:relatedwork}
Quantum switches are important components of quantum networks, and have attracted increasing research interest \cite{NaiVarGuhTow:20,VarGuhNaiTow:21,VarGuhNaiTow:20a,VarGuhNaiTow:20b}. In \cite{NaiVarGuhTow:20}, a quantum switch that serves multipartite entanglement to a set of end nodes is analyzed. Closed-form expressions for the expected number of stored qubits and the capacity are obtained. Moreover, the necessary and sufficient condition for the stability of the quantum switch is provided. In \cite{VarGuhNaiTow:21}, a similar setup that focuses on bipartite entanglement distribution is considered. Compared to \cite{NaiVarGuhTow:20}, the model of the quantum switch in \cite{VarGuhNaiTow:21} is more general, accounting for decoherence of quantum states in the memory and finite memory size. In \cite{VarGuhNaiTow:20b}, the quantum switch is assumed to have the capability of serving bipartite and tripartite entanglement to end nodes. A collection of randomized switching policies are evaluated, and it is shown that time-division multiplexing between bipartite and tripartite entanglement switching is not optimal. The setup in this manuscript is significantly different from these studies in three aspects. First, entanglement generation between the quantum switch and end nodes in these studies is formulated as a continuous-time Markov chain, i.e., at each time slot, one and only one of entanglement pair is generated through the quantum channels. In this manuscript, instead of the continuous-time Markov chain, we adopt the discrete-time Markov chain, which is shown to be much more challenging for analysis \cite{VarGuhNaiTow:20a}. Second, \cite{NaiVarGuhTow:20,VarGuhNaiTow:21,VarGuhNaiTow:20b,VarGuhNaiTow:20a} implicitly assume that the number of entanglement requests for every pair of users is infinite at any time slot, and when the quantum switch performs entanglement swapping or GHZ projection successfully, the generated bipartite or tripartite entanglement is immediately released from memory to address the entanglement requests. In this manuscript, the entanglement requests randomly arrive at the switch according to a stochastic process model. Correspondingly, the definition  of stability is different from \cite{NaiVarGuhTow:20}, and we focus on the unaddressed entanglement requests at the quantum switch. Third, one of the contributions of this manuscript is the design of entanglement swapping protocols, whereas in \cite{NaiVarGuhTow:20,VarGuhNaiTow:21,VarGuhNaiTow:20a,VarGuhNaiTow:20b}, the operations of the quantum switch are relatively simple. In particular, in \cite{NaiVarGuhTow:20,VarGuhNaiTow:21,VarGuhNaiTow:20a}, the quantum switch performs entanglement swapping or GHZ projection whenever there are sufficient entanglement pairs between the switch and end nodes. In \cite{VarGuhNaiTow:20b}, the switch needs to decide whether to perform entanglement swapping or GHZ projection, but does not need to choose  which two or three entangled pairs to operate on. The reason for such differences is the introduction of entanglement requests, and the objective of the switch is to address these requests instead of maximizing the entanglement switching rate.

Entanglement swapping protocols are proposed for networks with other structures than the star-shaped ones. In a recent paper \cite{VasTow:21}, entanglement distribution for a network consisting of quantum switches and users is considered. Similarly to this manuscript, entanglement requests are considered. However, the setup in this manuscript is significantly different that in \cite{VasTow:21}. First, the lifetime of a qubit is assumed to be one cycle in \cite{VasTow:21}, but the lifetime is assumed to be infinite in this manuscript. Second, the protocols proposed in \cite{VasTow:21} have high computational complexities, whereas most of the protocols, especially the on-demand protocols, proposed in this manuscript are efficient. Third, we use the NetSquid simulator in this manuscript, accounting for practical physical models. Another example of entanglement swapping protocols is \cite{ShchuSchLoo:19}, where the authors propose an approach to calculate the average waiting time for generating an entangled pair in quantum repeater chains. Such average waiting time depends on entanglement swapping schemes, and different schemes are compared in \cite{ShchuSchLoo:19} to minimize the average waiting time. Similarly to the studies on the quantum switch, this work aims at maximizing entanglement generating rate rather than addressing entanglement requests between end nodes. The maximum possible entanglement rate is also unclear if the number of nodes in repeater chains is more than five in the setup considered in \cite{ShchuSchLoo:19}.

\section{System Model}\label{sec:sys_model}


Consider a star-shape network consisting of $K+1$ nodes, where node $0$ is a quantum switch and the rest are end nodes. The quantum switch has $K$ interfaces that serve \ac{epr} pairs, where interface $k$ serves \ac{epr} pairs between the switch and node $k$, $k\in {\Set{K}} = \{1,2,\dotsc, K\}$. Time is slotted and at each time slot $t$, three types of events may occur, described as follows.


\renewcommand{\arraystretch}{1.5}
\begin{table*}[htbp]
\vspace{-0mm}
\centering
\begin{tabular}{  p{1.5cm}  | p{11.5cm} }
\hline
\textbf{Notation}& \textbf{Definition}\\
\hline
\rowcolor{blue!10!white} $K$& Number of end nodes \\
 $\mathcal{K}$ & Set of end nodes \\
\rowcolor{blue!10!white}$p_k$& Success probability of generating an \ac{epr} pair between the switch and node $k$ \\
$\ket{\Psi_{ij}}$ & \ac{epr} pair between node $i$ and $j$
\\
\rowcolor{blue!10!white}$q$& Success probability of entanglement swapping \\
 $\A_{ij}(t)$ & Number of entanglement requests between nodes $i$ and $j$ at time $t$\\
\rowcolor{blue!10!white}$\Ent_{ij}(t)$ & Number of $\ket{\Psi_{ij}}$ stored in nodes $i$ and $j$ at time $t$  \\
$\Ent_{0j}(t)$ & Number of $\ket{\Psi_{0j}}$ stored in the switch and node $j$ at time $t$  \\
\rowcolor{blue!10!white} $\UU_{ij}(t)$ & Number of requests for $\ket{\Psi_{ij}}$ at time $t$\\
 $\F_{ij}(t)$ & Number of $\ket{\Psi_{0i}}$ and $\ket{\Psi_{0j}}$ consumed to create $\ket{\Psi_{ij}}$ at time $t$ \\
\rowcolor{blue!10!white}$\BSN$ & Maximum number of entanglement swaps per time slot \\
$\RF_{ij}(t)$ & Number of successfully generated $\ket{\Psi_{ij}}$ at time $t$ via entanglement swapping\\
\rowcolor{blue!10!white}$\CC_{0i}$(t) & Number of successfully generated  $\ket{\Psi_{0i}}$ at time $t$ via the quantum channel
\\
$\lambda_{ij}$ & Rate of entanglement request $\A_{ij}(t)$ \\
\rowcolor{blue!10!white}$\M{H}(t)$ & All the history information at time $t$  \\
$A_\mathrm{max}$& Upper bound for the second moment of $\A_{ij}(t)$ defined in \eqref{eq:A_max_def}\\
\rowcolor{blue!10!white}$\M{\Lambda}$ & Capacity region 
\end{tabular}
\label{tab:resultsInference}
\vspace{3mm}\caption{Notations of Important Quantities.}
\end{table*}

\emph{Entanglement Generation:} The quantum switch attempts to generate \ac{epr} pairs with end nodes. An \ac{epr} pair between the quantum switch and node $k$ is generated with probability $p_k$, $k\in {\Set{K}}$ using a quantum channel. One qubit of each \ac{epr} pair is stored at the  switch and the other at the end node. Let $\CC_{0i}(t)$ denotes the number of \ac{epr} pairs $\ket{\Psi_{0i}}$ generated between the quantum switch and node $i\in \Set{K}$ at time slot $t$, and we assume that $\{\CC_{0i}(t):t\geq 0\}$, $i\in \Set{K}$ are mutually independent Bernoulli processes.\footnote{If multiplexing techniques can be used, $\{\CC_{0i}(t)\}$ can be modelled as Binomial random variables. Results in this work can be easily generalized to accommodate multiplexing techniques.}


\emph{Entanglement Swapping:} The quantum switch performs entanglement swapping operations. In particular, an \ac{epr} pair $\ket{\Psi_{ij}}$ is created with probability $q$ by consuming two \ac{epr} pairs, $\ket{\Psi_{0i}}$ and $\ket{\Psi_{0j}}$.\footnote{The entanglement swapping probability does not need to be the same for all node pairs. Results in this work can be easily generalized to accommodate different entanglement swapping probabilities among node pairs.}

\emph{Entanglement Request:} {During time slot $t$, entanglement requests randomly arrive at the switch, and the quantum switch maintains a queue for storing entanglement requests. Let $\A_{ij}(t)$ denote the number of entanglement requests between nodes $k$ and $j$ at time slot $t$, and we assume that $\{\A_{ij}(t):t\ge 0\}$ are mutually independent sequences of random variables.} 



For the entanglement requests $\{ \A_{ij}(t):t \geq 0\}$, we assume it is a stationary and ergodic process with rates $\lambda_{ij}$. Then the Birkhoff-Khinchin theorem \cite{Wal:B82} states that with probability one, 
\begin{align}\label{eq:BKtheorem}
\lim_{t\rightarrow \infty} \frac{1}{t}\sum_{\tau = 0}^{t-1} \A_{ij}(\tau) = \lambda_{ij}, \quad \forall i,j\in \Set{K}.
\end{align}
 
\subsection{System Dynamics}\label{sec:system_dynamics}
We now describe the variables and evolution of the switch. Let $\Ent_{ij}(t)$ denote the number of \ac{epr} pairs $\ket{\Psi_{ij}}$ stored in nodes $i,j\in \Set{K}$ at time $t\ge 0$. Let $\UU_{ij}(t)$ denote the number of pending entanglement requests for $\ket{\Psi_{ij}}$ at time $t\ge 0$, $i,j\in \Set{K}$. At each time slot, the quantum switch makes decisions about what link \ac{epr} pairs to perform entanglement swapping operations on. In particular, the quantum switch attempts to create  entanglement $\ket{\Psi_{ij}}$ by consuming $\F_{ij}(t)$ pairs of $\ket{\Psi_{0i}}$ and $\ket{\Psi_{0j}}$ from the stored entanglement in the quantum switch, $i, j\in \Set{K}$.\footnote{We do not distinguish between the order of nodes $i$ and $j$, i.e., $\F_{ij}(t) = \F_{ji}(t)$, $\forall i, j \in \Set{K}$.} The following constraints need to be satisfied:
\begin{align*}
\sum_{i\in \Set{K}}\F_{ij}(t) \leq  \Ent_{0j}(t) + \CC_{0j}(t),\quad \forall j\in {\Set{K}};t\ge 0.
\end{align*}
Moreover, we consider that at each time slot, the quantum switch can perform at most $\BSN$ entanglement swapping operations, leading to the following constraint:
\begin{align}\label{eq:ES_bound}
\sum_{ i,j \in \Set{K}}\F_{ij}(t) \leq \BSN;t \ge 0.
\end{align}
Outcomes of entanglement swapping operations are independent events, each succeeding with probability $q$. Let $\RF_{ij}(t)$ denote the number of successfully generated pairs $\ket{\Psi_{ij}}$, $i,j\in \Set{K}$. Let $[x]^+ = \max\{x, 0\}$. We assume entanglement swapping is performed at the beginning of each time slot, whereas the entanglement requests may arrive at any time during a time slot. Then, $\UU_{ij}(t)$ and $\Ent_{0i}(t)$ evolve as follows:
\begin{align*}
\UU_{ij}(t+1) &= [ \UU_{ij}(t)\Cb{+ \A_{ij}(t) }- \Ent_{ij}(t) -  \RF_{ij}(t)]^{+} ,\quad \forall i, j\in \Set{K} \\
\Ent_{ij}(t+1) & =  [ \Ent_{ij}(t) + \RF_{ij}(t) - \UU_{ij}(t)-\Cb{ \A_{ij}(t) }]^+,  \quad \forall i, j\in \Set{K}\\
\Ent_{0i}(t+1) & = \Ent_{0i}(t) - \sum_{j\in\Set{K}}\F_{ij}(t) + \CC_{0i}(t), \quad \forall i\in \Set{K}.
\end{align*}
Without loss of generality, we assume $\UU_{ij}(0) = \Ent_{ij}(0)=0$, $i, j\in \Set{K}$. Since no entanglement swapping is performed at time 0, we have $\RF_{ij}(0)=0$.

With the introduction of the system dynamics, we further assume that the second moment of entanglement requests is bounded at every time slot, regardless of history. Specifically, let
\begin{align}\label{eq:history}
\M{H}(t) = (\Ent_{ij}(\tau), \UU_{ij}(\tau), \A_{ij}(\tau), \RF_{ij}(\tau), \CC_{0i} , i, j\in {\Set{K}} )|_{\tau=0}^{t-1}
\end{align}
and for $i, j\in\Set{K}$ and $t$, we assume for any $\V{h}$ that is an instantiation of $\M{H} (t)$,
\begin{align}\label{eq:A_max_def}
\mathbb{E}\big[ {A}^2_{ij}(t)| \M{H}(t)=\V{h}\big] \leq A^2_\mathrm{max}, \quad \forall i, j\in\Set{K}.
\end{align}
 It follows that
 \begin{align}
 \mathbb{E}\big[  {A}_{ij}(t)| \M{H}(t)=\V{h}\big] &\leq A_\mathrm{max} \quad i, j\in\Set{K}\label{eq:A_average_bound}\\
 \mathbb{P}\big[	 {A}_{ij}(t)>A_\mathrm{max}+V| \M{H}(t)=\V{h}	\big]&\leq  \frac{A^2_\mathrm{max}}{V^2}\quad i, j\in{\Set{K}}, V>0 \label{eq:A_outage_bound}
 \end{align}
 where \eqref{eq:A_average_bound} follows because $\mathbb{E}[{X}] \leq \sqrt{\mathbb{E}[{X}^2]}$ and \eqref{eq:A_outage_bound} follows because of Chebyshev's inequality.

The goal of the quantum switch is to maintain as small a queue backlog $\UU_{ij}(t)$ for user pairs $i,j$ as possible and to stabilize the system. For now, we assume that there is no limit to the number of \ac{epr} pairs that can be stored in the nodes. Moreover, we assume that a qubit never decoheres. These assumptions will be relaxed in a later section of this manuscript.

\subsection{Stability and Capacity Region}
We follow the definition of stability in classical networks \cite{NeeModRoh:05}. Consider the following function:
\begin{align}\label{eq:g_function}
g_{ij}(V) := \limsup_{t\rightarrow \infty} \frac{1}{t}\sum_{\tau = 0}^{t-1} \mathbb{P}[\,{\UU}_{ij}(\tau) > V\,], \quad \forall i,j\in \Set{K}.
\end{align} 
This function characterizes the fraction of time that the number of unfinished requests for \ac{epr} pairs $\ket{\Psi_{ij}}$ exceeds a certain value $V$.

\begin{definition}
The quantum switch system is stable if $g_{ij}(V)\rightarrow 0$ as $V\rightarrow \infty$, for all $i, j\in \Set{K}$.
\end{definition}

We can then define the capacity region for the quantum switch system as follows.
\begin{definition}
The capacity region for the quantum switch system is the closure of the set of matrices $(\lambda_{ij})_{i,j\in \Set{K}}$ such that there exists an entanglement swapping algorithm that stabilizes the switch.
\end{definition}

In the following theorem, it is shown that the capacity region can be viewed as the set of all long-term entanglement requesting rates that the network can support.

\begin{theorem}[Capacity Region] \label{thm:capacity_region}
With given parameters $p_k$, $k\in \Set{K}$, $q$, and $\BSN$, the capacity region $\V{\Lambda}$ is the set of all matrices $(\lambda_{ij})_{i,j\in \Set{K}}$ for which there exist non-negative variables $\{f_{ij}\}_{i,j\in \Set{K}}$ satisfying:\footnote{We do not distinguish between the order of nodes $i$ and $j$, i.e., $f_{ij} = f_{ji}$, $\forall i, j\in\Set{K}$.}
\begin{align}
\sum_{i\in \Set{K}} f_{ij} &\leq p_j, \quad \forall j\in \Set{K} \label{eq:capacity_necessary_1} \\
\sum_{i,j\in \Set{K}} f_{ij} &\leq \BSN \label{eq:capacity_necessary_2} \\
\lambda_{ij}& \leq q\, f_{ij}, \quad \forall i,j\in \Set{K}. \label{eq:capacity_necessary_3}
\end{align}
\end{theorem}
The physical meaning of $f_{ij}$ is the rate of entanglement swapping that consumes $\ket{\Psi_{0i}}$  $\ket{\Psi_{0j}}$, referred to as entanglement flow variables. The equaility \eqref{eq:capacity_necessary_1} constrain the flow variables so that their sum is no greater than the "raw" entanglement rate; the inequality \eqref{eq:capacity_necessary_1} corresponds to the constraint \eqref{eq:ES_bound}; and \eqref{eq:capacity_necessary_3} corresponds to the definition of $q$. The proof of Theorem \ref{thm:capacity_region} consists of the necessity part (details can be found in Appendix \ref{apd:proof_capacity_region}) and the sufficiency part, shown in the next two sections where protocols are designed and shown to stabilize the quantum switch whenever $(\lambda_{ij})$ is an interior point of $\V{\Lambda}$.

\section{Protocol Design}\label{sec:protocol_design}
In this section, we design the stationary and the on-demand protocols for entanglement swapping in a quantum switch. More importantly, we show that these protocols stabilize the switch when the rate matrix is within the capacity region. For simplicity, we assume that the quantum switch can perform an arbitrary number of entanglement swapping operations (i.e., $W=\infty$), but all of the protocols can be easily modified to satisfy the \Cb{constraint on} the number of entanglement swapping operations \Cb{per time slot}. \Cb{The stationary protocol requires knowing the rate matrix and channel parameters, and it does not depend on the states of the system. The on-demand protocols do not require any knowledge of rate matrix or channel parameter. Though they need to solve an optimization problem every time slot, the optimization problem can be efficiently solved.}


\subsection{Stationary Protocol}
We first design a stationary protocol $\pi_\mathrm{STAT}$. Suppose the rate matrix $(\lambda_{ij})_{i, j\in \Set{K}}$ is known to the quantum switch and there exists $\epsilon>0$ such that 
\begin{align}\label{eq:lambda_in_region}
(\lambda_{ij}+\epsilon)_{i,j\in\Set{K}} \in \V{\Lambda}.
\end{align} 
Then there exists a set of variables $\{\tilde{f}_{ij}\}_{i,j\in \Set{K}}$ such that
\begin{align}
\sum_{i\in \Set{K}} \tilde{f}_{ij} &\leq p_j, \quad \forall j\in \Set{K} \label{eq:sta_tilde_f1} \\
\lambda_{ij} +\epsilon &\leq q \,\tilde{f}_{ij}, \quad \forall i,j\in \Set{K}.\label{eq:sta_tilde_f2} 
\end{align}
In fact, determining $\{\tilde{f}_{ij}\}_{i, j\in \Set{K}}$ is straightforward since we can set $\tilde{f}_{ij} = (\lambda_{ij}+\epsilon)/q$. For an \ac{epr} pair $\ket{\Psi_{0i}}$ generated at time slot $t$, $i\in \Set{K}$, label it as $(i, j)$ with probability $\tilde{f}_{ij}/p_i$. Let ${\Set{M}}^{i}_{ij}(t)$ denote the set of \ac{epr} pairs $\ket{\Psi_{0i}}$ labelled as $(i, j)$ that are not consumed for entanglement swapping up to time slot $t$. If ${\Set{M}}_{ij}^{i}(t)\neq \varnothing$ and ${\Set{M}}_{ij}^{j}(t)\neq \varnothing$, the quantum switch performs entanglement swapping to create \ac{epr} pair $\ket{\Psi_{ij}}$ by consuming \ac{epr} pairs $\ket{\Psi_{0i}}$ and $\ket{\Psi_{0j}}$ until either ${\Set{M}}_{ij}^{i}(t)$ or ${\Set{M}}_{ij}^{j}(t)$ is empty.

The stationary protocol requires knowing not only the parameters $p_k$, $k\in \Set{K}$, but also the rate matrix $(\lambda_{ij})_{i,j\in \Set{K}}$ in order to find $\{\tilde{f}_{ij}\}_{i,j \in \Set{K}}$. Such knowledge of all the rates may not be available in practice, and we may need protocols that requires less information. This motivates the design of \Cb{on-demand protocols}.

\subsection{On-demand Protocols}
The on-demand protocols are in fact  a family of protocols. They have the flexibility to prioritize entanglement requests as long as  certain constraints are satisfied. Such flexibility makes on-demand protocols computationally efficient.


\emph{On-demand Protocols:} At each time slot, the quantum switch attempts to create entanglement $\ket{\Psi_{ij}}$ using $\F_{ij}$ pairs of entanglement $\ket{\Psi_{0i}}$ and $\F_{ij}$ pairs of entanglement $\ket{\Psi_{0j}}$. The decisions $\{\F_{ij}\}_{i,j\in\Set{K}}$ need to satisfy the following constraints:
\begin{align}\label{eq:entanglement_constraint}
&\sum_{i\in\Set{K}} \F_{ij} \leq \Ent_{0j}(t)+\CC_{0j}(t), \quad j\in\Set{K} \\ \label{eq:demand_constraint}
&\F_{ij}  \leq \UU_{ij}(t)+\A_{ij}(t), \quad i,j\in\Set{K} \\ \nonumber
&  \F_{ij}=\F_{ji} \in \mathbb{N},\quad i, j\in \Set{K} \\ \label{eq:equality_constraint}
&\Big(\Ent_{0i}(t) +\CC_{0i}(t)- \sum_{k\in\Set{K}} \F_{ik}\Big)\Big(\Ent_{0j}(t) +\CC_{0j}(t)- \sum_{k\in\Set{K}} \F_{kj}\Big)\big(\UU_{ij}(t)+\A_{ij}(t)- \F_{ij}\big) = 0,\quad i, j\in\Set{K}.
\end{align}
Then the quantum switch attempts to create entanglement $\ket{\Psi_{ij}}$ using $\F_{ij}$ pairs of entanglement $\ket{\Psi_{0i}}$ and $\F_{ij}$ pairs of entanglement $\ket{\Psi_{0j}}$. An on-demand protocol is denoted by $\pi_{\mathrm{od}}$.

The intuition of \eqref{eq:demand_constraint} in on-demand protocols is that the quantum switch creates entanglement to address the existing request of entanglement, but not excess entanglement for future entanglement request. Moreover, the quantum switch should not ``waste'' any opportunities to create entanglement, in the sense that for any $i, j$, it should attempt to create as many pairs of $\ket{\Psi_{ij}}$ as possible provided that \eqref{eq:demand_constraint} holds. This gives the condition \eqref{eq:equality_constraint}. Note that there may be multiple choices of $\{\F_{ij}\}_{i,j\in \Set{K}}$ that satisfy the constraints. The quantum switch can select any one of them for entanglement swapping.

One way to satisfy the constraints \eqref{eq:entanglement_constraint}-\eqref{eq:equality_constraint} is to first set $F_{ij}=0$, $i, j\in\Set{K}$ and then check the unfinished entanglement requests $\{\UU_{ij}(t)\}_{i,j\in\Set{K}}$ in an arbitrary order. Given $i,j$, if $\UU_{ij}(t)>0$, we set $F_{ij}$ to be as large as possible provided that \eqref{eq:entanglement_constraint} and \eqref{eq:demand_constraint} hold. After an iteration of all $i, j$ in $\Set{K}$, we obtain a desirable solution. Note that the complexity is $O(K^2)$ since the iteration is over all $i, j\in\Set{K}$. Comparatively, the stationary protocol also has a complexity $O(K^2)$ in average since the switch has to randomly label every \ac{epr} generated between the switch and nodes. 

\section{Stability Analysis}
In this section, we show the protocols designed in Section \ref{sec:protocol_design}  stabilize the quantum switch whenever $(\lambda_{ij})$ is an interior point of $\V{\Lambda}$.

\subsection{Stationary Protocol}\label{sec:stability_proof_stationary}
The proof of the stability of the stationary protocol is organized in four steps.

\textbf{Step 1}: Introduce an auxiliary protocol $\pi_{\mathrm{STAT-DISCARD}}$.

To show that the stationary protocol stabilizes the quantum switch system, we consider an auxiliary protocol $\pi_{\mathrm{STAT-DISCARD}}$, which is the same as the stationary protocol $\pi_{\mathrm{STAT}}$ except that it periodically discards used entanglement pairs. The length of the period, denoted by $T_0$, is chosen
to be sufficiently large such that the average number of entanglement request and the average number of \ac{epr} pairs between the quantum switch and end nodes are close enough to the expected value. The reason we want to introduce such an auxiliary protocol that discards entanglement pairs is to bound $\RF_{ij}(t)$, $i,j\in\Set{K}$, which is crucial in the stability analysis.

\emph{Stationary Discard Protocol:} Set  $\delta$ to be a sufficiently small positive number such that
\begin{align}\label{eq:delta_condition}
\lambda_{ij}+\delta+{\epsilon}/{4} \leq q(\tilde{f}_{ij}-\epsilon/2)(1-\delta).
\end{align}
Such a $\delta$ must exist since as $\delta$ goes to 0, the left hand side goes to $\lambda_{ij}+{\epsilon}/{4}$ and the right hand side goes to $q(\tilde{f}_{ij}-\epsilon/2)$, which is greater than $\lambda_{ij}+{\epsilon}/{2}$ because of \eqref{eq:sta_tilde_f2}. Consider a positive integer $T_0$ such that for any time slot $t_0$ and any $\V{h}$ that is an instantiation of $\M{H}(t_0)$, the following conditions hold:
\begin{align}\label{eq:ergodic_A}
\Big|\lambda_{ij} - \frac{1}{T_0}\sum_{\tau = t_0}^{t_0+T_0-1}\mathbb{E}\big[ {\A}_{ij}(\tau)| \M{H}(t_0)=\V{h}\big]\Big | &\leq \delta, \quad \forall i, j\in\Set{K} \\
\mathbb{P}\bigg[X_{ij}< T_0 ( \tilde{f}_{ij}-\epsilon/2)\bigg]&\leq \delta, \quad \forall i, j\in\Set{K} \label{eq:min_binomial_converge}
\end{align}
where $\M{H}(\cdot)$ represents the history, $X_{ij} = \min\Big\{\sum_{\tau = t_0}^{t_0+T_0-1} {X}^{(i)}_{ij}(\tau),\sum_{\tau = t_0}^{t_0+T_0-1} {X}^{(j)}_{ij}(\tau) \Big\}$, in which $X^{(i)}_{ij}(\tau)$ and ${X}^{(j)}_{ij}(\tau)$ are i.i.d Bernoulli random variables with the same mean $\tilde{f}_{ij}$ defined in \eqref{eq:sta_tilde_f1} and \eqref{eq:sta_tilde_f2}. The variables  $X^{(i)}_{ij}(\tau)$ (${X}^{(j)}_{ij}(\tau)$) can be viewed as the number of \ac{epr} pairs between the switch and end node $i$ ($j$) generated at time $\tau$ and labelled as $(i, j)$; correspondingly, the random variable $X_{ij}$ is the number of total attempts for distributing $\ket{\Psi_{ij}}$ using \ac{epr} pairs generated during $T_0$ time slots. For \eqref{eq:ergodic_A}, such a value $T_0$ must exist because $\A_{ij}(t)$ is stationary and ergodic; for \eqref{eq:min_binomial_converge}, the existence of such a value $T_0$ follows from the application of a Chernoff bound to ${X}_{ij}(\tau)$ and ${Y}_{ij}(\tau)$. The auxiliary protocol $\pi_{\mathrm{STAT-DISCARD}}$ performs the same operation as $\pi_{\mathrm{STAT}}$ except that every $T_0$ slots (in particular, at time $t=mT_0-1$, $m=1,2,3\dotsc$), the unused \ac{epr} pairs $\ket{\Psi_{ij}}$, $i, j\in\Set{K}$ and $\ket{\Psi_{0i}}$, $i\in\Set{K}$ are discarded. Evidently if $\pi_{\mathrm{STAT-DISCARD}}$ can stabilize the quantum switch system, then $\pi_{\mathrm{STAT}}$ can too. 


\textbf{Step 2}: Introduce a $T$-step Lyapunov function and the connection between this function and stability of a quantun switch.

Define a Lyapunov function of unprocessed entanglement request a follows
\begin{align}\label{eq:L}
L(\M{u}) = \sum_{i, j\in\Set{K}} (u_{i j})^2,\quad \V{u}\in \mathbb{N}^{K\times K}.
\end{align}
 For a given protocol $\pi$, consider the following $T$-step Lyapunov drift 
\begin{align}\label{eq:delta_pi_T}
\Delta^{\pi}_{T}\big(\M{u}, t\big) := \mathbb{E}\big[ L(\M{U}(t+T)) - L(\M{u}) | \M{U}(t)= \M{u}\big].
\end{align}
\begin{lemma}\label{lemma:T_step_Lyapunov_stable}
If there exists a positive integer $T$ such that for time slots $t\in \{mT+J: m = 0,1,2,3,\dotsc \}$, where $J\in\{0,1,2,\dotsc, T-1\}$, the $T$-step Lyapunov drift using a protocol $\pi$ satisfies
\begin{align}\label{eq:condition_T_LD}
\Delta^{\pi}_{T}\big(\M{u},t\big) \leq C - \sum_{i,j\in\Set{K}} \theta_{ij} u_{i j}.
\end{align}
for some positive constants $C$, $\{\theta_{ij}\}$, where $\M{u}\in\mathbb{N}^{K\times K}$ and $u_{i j}$ is the $ij$th element of $\M{u}$, and if $\mathbb{E}[L(\M{U}(j))]<\infty$, $j=0,1,2,\dotsc, J$, then the quantum switch is stable.
\end{lemma}
\begin{IEEEproof}
See Appendix \ref{apd:T_step_Lyapunov_stable}.
\end{IEEEproof}

\begin{lemma}\label{lemma:T_step_lyapunov_bound}
For any control policy $\pi$, the $T$-step Lyapunov drift at any slot $t_0 $ satisfies
\begin{align*}
\Delta^{\pi}_{T}\big(\M{u},t_0\big) & \leq \sum_{i,j\in\Set{K}}
 \mathbb{E}\bigg[ \Big( \sum_{t = t_0}^{t_0+T-1}{\RF}_{ij}(t)\Big)^2 \big | \M{U}(t_0)=\M{u} \bigg] + \frac{(KTA_\mathrm{max})^2}{2}\\
 &\quad+\sum_{i,j\in\Set{K}} 2u_{ij}
 \mathbb{E}\Big[\sum_{t = t_0}^{t_0+T-1}{\A}_{ij}(t) - {\RF}_{ij}(t) \big | \M{U}(t_0)=\M{u} \Big].
\end{align*}
\end{lemma}
\begin{IEEEproof}
See Appendix \ref{apd:T_step_Lyapunov_bound}.
\end{IEEEproof}

Note that Lemmas \ref{lemma:T_step_Lyapunov_stable} and \ref{lemma:T_step_lyapunov_bound} hold for any protocol, \Cb{including} the stationary protocol.

\textbf{Step 3}: Provide an upper bound for the $T$-step Lyapunov function of the  stationary discard protocol $\pi_{\mathrm{STAT-DISCARD}}$.

We consider the $T_0$-step Lyapunov drift $\Delta^{\pi_{\mathrm{STAT-DISCARD}}}_{T_0}\big(\M{u},t_0\big)$ for $t_0=mT$, $m\in\{0,1,2,\dotsc\}$. Using Lemma \ref{lemma:T_step_lyapunov_bound}, we have
\begin{align*}
\Delta^{\pi_{\mathrm{STAT-DISCARD}}}_{T_0}\big(\M{u},t_0\big) & \leq \sum_{i,j\in\Set{K}}
 \mathbb{E}\Big[ \Big( \sum_{\tau = t_0}^{t_0+T_0-1}{\RF}_{ij}(\tau)\Big)^2 \big | \M{U}(t_0)=\V{u} \Big] + \frac{(KT_0A_\mathrm{max})^2}{2}\\
 &\quad+\sum_{i,j\in\Set{K}} 2u_{ij}
 \mathbb{E}\Big[\sum_{\tau = t_0}^{t_0+T_0-1}{\A}_{ij}(\tau) - {\RF}_{ij}(\tau) \big |  \M{U}(t_0)=\V{u}\Big].
\end{align*}
We then bound each term on the right hand side of the expression above. Since $\pi_{\mathrm{STAT-DISCARD}}$ discards the unused \ac{epr} pair $\ket{\Psi_{ij}}$, $i, j\in\Set{K}$ and $\ket{\Psi_{0i}}$, $i\in\Set{K}$ every $T_0$ slots,
\begin{align}\label{eq:stat_DISCARD_bound_1}
\sum_{\tau= t_0}^{t_0+T_0-1}\RF_{ij}(\tau) \leq \sum_{\tau= t_0}^{t_0+T_0-1}\F_{ij}(\tau) \leq \sum_{\tau= t_0}^{t_0+T_0-1}\CC_{0i}(\tau) \leq T_0.
\end{align}
We also have
\begin{align}\label{eq:E_A_bound}
 \mathbb{E}\Big[\sum_{\tau = t_0}^{t_0+T_0-1}{\A}_{ij}(\tau)  \big | \M{U}(t_0)=\M{u} \Big] &\leq T_0( \lambda_{ij}+\delta) \\ 
\mathbb{E}\Big[\sum_{\tau = t_0}^{t_0+T_0-1}{\RF}_{ij}(\tau)  \big | \M{U}(t_0) =\M{u}\Big] &= \mathbb{E}\Big[\sum_{\tau = t_0}^{t_0+T_0-1}{\RF}_{ij}(\tau)  \Big] = q \,\mathbb{E}\Big[\sum_{\tau = t_0}^{t_0+T_0-1}{\F}_{ij}(\tau)  \Big]  \label{eq:Wald}\\
&\geq q T_0(\tilde{f}_{ij}-\epsilon/2)\mathbb{P}\Big[ \sum_{\tau = t_0}^{t_0+T_0-1}{\F}_{ij}(\tau)  \geq T_0(\tilde{f}_{ij}-\epsilon/2)\Big] \nonumber\\
& \geq qT_0(\tilde{f}_{ij}-\epsilon/2)(1-\delta) \label{eq:E_mu_bound}
\end{align}
where \eqref{eq:E_A_bound} is due to \eqref{eq:ergodic_A}, \eqref{eq:Wald}
is because ${\RF}_{ij}(\tau)$ represents the results of ${\RF}_{ij}(\tau)$ swaps and ${\RF}_{ij}(\tau)$ is independent of $\M{U}(t_0)$ in the stationary discard protocol, and \eqref{eq:E_mu_bound} is due to \eqref{eq:min_binomial_converge}. Recall that $\delta$ is selected such that \eqref{eq:delta_condition} holds, and we have
\begin{align}\label{eq:stationary_delta_bound}
\sum_{i,j\in\Set{K}} u_{ij}
 \mathbb{E}\Big[\sum_{\tau = t_0}^{t_0+T_0-1}{\A}_{ij}(\tau) - {\RF}_{ij}(\tau) \big | \M{U}(t_0)=\V{u} \Big] \leq -\frac{\epsilon T_0}{4}\sum_{i,j\in\Set{K}} u_{ij}.
\end{align}
Together with \eqref{eq:stat_DISCARD_bound_1}, we have
\begin{align*}
\Delta^{\pi_{\mathrm{STAT-DISCARD}}}_{T_0}\big(\M{u},t_0\big) & \leq \frac{(KT_0)^2}{2}+ \frac{(KT_0A_\mathrm{max})^2}{2}-\frac{\epsilon T_0}{4}\sum_{i,j\in\Set{K}} u_{ij}.
\end{align*}

\textbf{Step 4}: Show the stability of $\pi_{\mathrm{STAT-DISCARD}}$ and $\pi_{\mathrm{STAT}}$.

Using Lemma \ref{lemma:T_step_Lyapunov_stable} and the inequality at the end of Step 3, we show that the quantum switch is stable. As a corollary, we show the following theorem.
\begin{theorem}
The quantum switch system is stable using the stationary protocol $\pi_{\mathrm{STAT}}$  if $(\lambda_{ij})_{i,j\in\Set{K}}$ is an interior point in $\V{\Lambda}$, i.e., \eqref{eq:lambda_in_region} holds for some $\epsilon>0$.
\end{theorem}

\subsection{On-demand Protocols}

We make the following assumption under which we can show the stability of the switch with on-demand protocols.

Assumption 1: For an arbitrary $\epsilon_0>0$, consider the following event:
\begin{align}
B_0:\exists i, j,~\bigg|\frac{1}{t+1}\sum_{\tau = 0}^{t}\A_{ij}(\tau)-\lambda_{ij}\bigg|>\epsilon_0.
\end{align}
the entanglement requests $\{\A_{ij}(t)\}_{i,j\in\Set{K}}$ satisfies
\begin{align}\label{eq:assumption_Aij}
\mathbb{E}\Big[\sum_{\tau = 0}^{t}{\A}_{ij}(\tau) \big | B_0\Big] \mathbb{P}[B_0]\leq c_1(\epsilon_0)
\end{align}
where $c_1(\epsilon_0)$ is a function of $\epsilon_0$ independent of $t$.

Note that Assumption 1 holds for many random processes. For example, if $\A_{ij}(t)$ is i.i.d over time and $\mathbb{V}\mathrm{ar}\Big[ \A_{ij}(t)\Big] = \sigma^2$, where $\sigma$ is a constant irrelevant of $t$, then one can use  Chebyshev's inequality to verify that Assumption 1 holds.


The proof of the stability of the on-demand protocols is organized in six steps. 

\textbf{Step 1}: Introduce the uncontidional 1-step Lyapunov drift of an on-demand protocol.

 Consider $\M{u}$ as a matrix with $ij$th element being $u_{ij}$. Define 
\begin{align}\label{eq:Lod}
L_{\mathrm{od}}(\M{u}) = \sum_{i,j\in\Set{K}} u_{ij}
\end{align}
as a Lyapunov function of unprocessed entanglement request. For a control policy, we consider the following unconditional 1-step Lyapunov drift
\begin{align}\label{eq:1_step_L}
\tilde{\Delta}^{\pi}_{1}(t) = \mathbb{E}\big[ L_{\mathrm{od}}(\M{U}(t+1)) - L_{\mathrm{od}}(\M{U}(t)) \big].
\end{align}
If we can provide an upper bound  \eqref{eq:1_step_L}, then we use the similar connection between the Lyapunov drift and the stability as in Section \ref{sec:stability_proof_stationary}. For any random event $S(t)$, we can rewrite $\tilde{\Delta}^{\pi}_{1}(t)$ as 
\begin{align}\label{eq:delta_pi_2}
\tilde{\Delta}^{\pi}_{1}(t)
& = \mathbb{P}\big [S(t) \big] \mathbb{E}\Big[ L_{\mathrm{od}}(\M{U}(t+1)) - L_{\mathrm{od}}(\M{U}(t)) | S(t)\Big] \\
&\quad +\mathbb{P}\big [\overline{S}(t) \big] \mathbb{E}\Big[ L_{\mathrm{od}}(\M{U}(t+1)) - L_{\mathrm{od}}(\M{U}(t))| \overline{S}(t)\Big].\label{eq:delta_pi_1}
\end{align}
We then choose a random event $S(t)$ for the derivation of an upper bound of $\tilde{\Delta}^{\pi}_{1}(t) $.

\textbf{Step 2}: Introduce a random event $S(t)$.

We consider a random event ${S}(t)$ defined as follows:
\begin{align}\label{eq:def_S}
{S}(t) = \Big\{ \sum_{i\in\Set{K}}	\big[{\UU}_{ij}(t)+\A_{ij}(t)\big]\leq {\Ent}_{0j}	(t)+\CC_{0j}(t), \forall j\in\Set{K}\Big\}
\end{align}
where ${\UU}_{ij}(t)$ and ${\Ent}_{0j}(t)$ are the number of backlog of entanglement request $\ket{\Psi_{ij}}$ and the number of entanglement $\ket{\Psi_{0j}}$ achieved by the used on-demand protocol. If the event ${S}(t)$ occurs, then the quantum switch has sufficient entanglement to address the entanglement request, i.e., $\F_{ij}(t) = \UU_{ij}(t)+\A_{ij}(t)$, $\forall i,j\in\Set{K}$ if an on-demand protocol is used.

\textbf{Step 3}: Provide an upper bound for \eqref{eq:delta_pi_2}.

As mentioned above, if ${S}(t)$ occurs, $\F_{ij} (t)= \UU_{ij}(t)+\A_{ij}(t)$, $\forall i,j\in\Set{K}$ and then
\begin{align}
\mathbb{P}\big [S(t)\big]\mathbb{E}\Big[{\UU}_{ij}(t+1)-{\UU}_{ij}(t) | S(t)\Big] \leq q\mathbb{P}\big [S(t) \big]\bigg[\mathbb{E}\Big[A_{ij}(t)|S(t)\Big]-\,\mathbb{E}\Big[ {\UU}_{ij}(t)|S(t)\Big] \bigg] \label{eq:S_01}.
\end{align}
Summing over all $i, j\in \Set{K}$ gives the desired upper bound. 

\textbf{Step 4}: Provide an upper bound for \eqref{eq:delta_pi_1}.

Evidently, $\UU_{ij}(t+1)-{\UU}_{ij}(t) \leq A_{ij}(t)$, $\forall i, j, t$. Then
\begin{align}
\mathbb{P}\big[\overline{S}\big]\mathbb{E}\Big[{\UU}_{ij}(t+1)-{\UU}_{ij}(t) | \overline{S}(t)\Big]  \leq  \mathbb{P}\big[\overline{S}(t)\big]\mathbb{E}\Big[A_{ij}(t)|\overline{S}(t)\Big]. \label{eq:S_00}
\end{align}

\textbf{Step 5}: Provide an upper bound for $\tilde{\Delta}^{\pi}_{1}(t)$.

Steps 3 and 4 give
\begin{align}\nonumber
\tilde{\Delta}^{\pi}_{1}(t) &\leq \sum_{i,j\in\Set{K}}\bigg[\lambda_{ij}-q\mathbb{P}[S(t)] \,\mathbb{E}\Big[ {\UU}_{ij}(t)|S(t)\Big] \bigg] \\ \nonumber
&= \sum_{i,j\in\Set{K}}\bigg[\lambda_{ij}-q \mathbb{E}\Big[ {\UU}_{ij}(t) \Big]  +q  \mathbb{P}[\overline{S}(t)] \mathbb{E}\Big[ \UU_{ij}(t)|\overline{S}(t)\Big] \bigg] \\
& \leq \sum_{i,j\in\Set{K}}\bigg[\lambda_{ij}-q \mathbb{E}\Big[ {\UU}_{ij}(t) \Big]  + q \mathbb{P}[\overline{S}(t)] \sum_{\tau=0}^{t}\mathbb{E}\Big[ \A_{ij}(\tau)|\overline{S}(t)\Big]\bigg]. \label{eq:delta_1_pi_bound}
\end{align}
The following lemma is used to upper bound \eqref{eq:delta_1_pi_bound}.
\begin{lemma}\label{lemma:lemma_long}
Under Assumption 1, if an on-demand protocol is used,
then there exists a constant $ c_2$ irrelevant of $t$, such that
for any $\tilde{i}$, $\tilde{j}\in\Set{K}$
\begin{align}\label{eq:cpnditional_exp_A}
\mathbb{E}\Big[ \sum_{\tau = 0}^{t}\A_{\tilde{i}\tilde{j}}(\tau)|\overline{S}(t)\Big]\mathbb{P}\big[\overline{S}(t)\big] \leq c_2.
\end{align}
\end{lemma}
The proof of Lemma \ref{lemma:lemma_long} can be found in Appendix \ref{apd:lemma_long}. Combining \eqref{eq:delta_1_pi_bound} and Lemma \ref{lemma:lemma_long}, we have
\begin{align}\label{eq:Lyapunov_upper_bound_od}
\tilde{\Delta}^{\pi}_{1}(t) &\leq\sum_{i,j\in\Set{K}}\Big[ \lambda_{ij}+qc_2-q \,\mathbb{E}\big[ {\UU}_{ij}(t)\big] \Big].
\end{align}

\textbf{Step 6}: Stability of the on-demand protocols.

Summing over $t$ from $t=0$ to $T-1$ at both sides of \eqref{eq:Lyapunov_upper_bound_od} gives
\begin{align*}
\frac{1}{T}\sum_{t=0}^{T-1}\sum_{i,j\in\Set{K}}q\mathbb{E}\big[\UU_{ij}(t)\big]\leq \sum_{i,j\in\Set{K}}(\lambda_{ij}+qc_2) + \sum_{i,j\in\Set{K}}\mathbb{E}\big[\UU_{ij}(0)\big]/T.
\end{align*}
Then for any $i_0,j_0\in\Set{K}$, for $V>0$,
\begin{align*}
\frac{1}{T}\sum_{t=0}^{T-1} \mathbb{P}\big[\UU_{i_0j_0}(t)>V\big]&\leq \frac{1}{VT}\sum_{t=0}^{T-1}\mathbb{E}\big[\UU_{i_0j_0}(t)\big] \leq \frac{\sum_{i,j\in\Set{K}}(\lambda_{ij}+qc_2)}{qV} + \frac{\sum_{i,j\in\Set{K}}\mathbb{E}\big[\UU_{ij}(0)\big]}{qTV}.
\end{align*}
Taking $\limsup_{t\rightarrow\infty}$, we see that $g_{i_0j_0}(V)$ defined in \eqref{eq:g_function} is on the order of $O(1/V)$. Taking limits as $V\rightarrow\infty$, we have $g_{ij}(V)\rightarrow 0$, $i,j\in\Set{K}$. We then prove the following theorem.
\begin{theorem}\label{thm:on_demand_general}
The quantum switch system is stable using an on-demand protocol $\pi_{\mathrm{od}}$ if $(\lambda_{ij})_{i,j\in\Set{K}}$ is an interior point in $\V{\Lambda}$ under Assumption 1.
\end{theorem}



\section{Zero Average Latency}

In previous sections, we show that the stationary protocol and the on-demand protocol can stabilize the switch if the rate matrix is an interior point of the capacity region. In this section, we take a step further and show that with slight modification, these protocols have zero average latency.

In fact, Little law implies that a protocol achieves zero average latency if for all $i, j\in\Set{K}$
\begin{align}\label{eq:zero_latency}
\limsup_{t\rightarrow \infty}\frac{1}{t}\sum_{\tau=0}^{t-1}\mathbb{E}[\UU_{ij}(\tau)]=0.
\end{align}
Therefore, we only need to show the expected queue length $\mathbb{E}[\UU_{ij}(t)]$ converges to 0. To do this, we need an assumption of the entanglement requests stronger than Assumption 1.

Assumption 2: For an arbitrary $\epsilon_0>0$, consider the following event:
\begin{align}
C_0:\exists i, j,~ \frac{1}{t+1}\sum_{\tau = 0}^{t}\A_{ij}(\tau) >\lambda_{ij}+\epsilon_0.
\end{align}
the entanglement requests $\{\A_{ij}(t)\}_{i,j\in\Set{K}}$ satisfies
\begin{align}\label{eq:assumption_Aij_virtual}
\lim_{t\rightarrow \infty}\mathbb{E}\Big[\sum_{\tau = 0}^{t}{\A}_{ij}(\tau) \big | C_0\Big] \mathbb{P}[C_0] = 0.
\end{align}

Note that though stronger than Assumption 1, Assumption 2 still holds for many random processes. For example, Assumption 2 holds if $\A_{ij}(t)$ is i.i.d over time, its support has an upper bound, and $\mathbb{P}\big[\sum_{\tau=0}^{t}\A_{ij}(t)>(t+1)(\lambda_{ij}+\epsilon_0)\big]\sim o(1/t)$.

\subsection{Zero Average Latency of Stationary Protocols}
With Assumption 2, we can show that the stationary protocol achieves zero average latency.

\begin{theorem}\label{thm:stationary_zero_latency}
The quantum switch system achieves zero average latency using the stationary protocol  if $(\lambda_{ij})_{i,j\in\Set{K}}$ is an interior point in $\V{\Lambda}$ under Assumption 2.
\end{theorem}
\begin{IEEEproof}
For  $\epsilon $ in \eqref{eq:lambda_in_region}, define events:
\begin{align}
&\widetilde{C}_0:  \frac{1}{t}\sum_{\tau = 0}^{t-1}\A_{ij}(\tau) >\lambda_{ij}+\epsilon/2 \\
&\widetilde{C}_1: \frac{1}{t}\sum_{\tau=0}^{t-1} \RF_{ij}(\tau)\leq  \lambda_{ij}+\epsilon/2 .
\end{align}

Note that if neither $\widetilde{C}_0$ nor $\widetilde{C}_1$ occurs, then $\Big( \sum_{\tau = 0}^{t-1} \A_{ij}(\tau) -\sum_{\tau = 0}^{t-1} \RF_{ij}(\tau)  \Big)^+=0$. As a consequence,
\begin{align}
\mathbb{E}\big[\UU_{ij}(t)\big] =\leq \mathbb{E}\Big[   \sum_{\tau = 0}^{t-1} \A_{ij}(\tau)  | \widetilde{C}_0  \Big]\mathbb{P}[\widetilde{C}_0] +  \mathbb{E}\Big[   \sum_{\tau = 0}^{t-1} \A_{ij}(\tau)  | \widetilde{C}_1  \Big]\mathbb{P}[\widetilde{C}_1] .\label{eq:two_terms_Uij}
\end{align}
The first term in \eqref{eq:two_terms_Uij} converges to zero due to Assumption 2. Regarding the second term in \eqref{eq:two_terms_Uij},
\begin{align*}
 \mathbb{E}\Big[   \sum_{\tau = 0}^{t-1} \A_{ij}(\tau)  | \widetilde{C}_1  \Big]\mathbb{P}[\widetilde{C}_1] =  \mathbb{E}\Big[   \sum_{\tau = 0}^{t-1} \A_{ij}(\tau)   \Big]\mathbb{P}[\widetilde{C}_1] =\lambda_{ij}t\mathbb{P}[\widetilde{C}_1].
\end{align*}
Using Chernoff bounds on $\sum_{\tau=0}^{t-1}\F_{ij}(\tau)$, $\sum_{\tau=0}^{t-1} X_{ij}^{(i)}$, and $\sum_{\tau=0}^{t-1} X_{ij}^{(j)}$, one can easily verify that $\mathbb{P}[\widetilde{C}_1]$ decays exponentially with respect to $t$. Therefore, \eqref{eq:two_terms_Uij} converges to zero as $t$ goes to infinity. This finishes the proof.
\end{IEEEproof}

\subsection{Virtual Requests}
As mentioned in the previous section, the stationary protocol requires the knowing the rate matrix $(\lambda_{ij})_{i,j\in\Set{K}}$, which may be available in practice. To address this issue, we develop the on-demand protocols with \emph{virtual requests}. These protocols are similar to the on-demand protocols except that we add some virtual requests to the system. In this way, the switch can create end-to-end entanglement before the requests come, so the latency can be made arbitrarily small. In particular, select $\alpha\in(1/2, 1)$ and define
\begin{align*}
\WA_{ij}(t) = \A_{ij}+\lceil (t+1)^{\alpha} \rceil - \lceil t^{\alpha} \rceil
\end{align*}
where the term $\lceil (t+1)^{\alpha} \rceil - \lceil t^{\alpha} \rceil$ is the virtual request. In this section, we also define
\begin{align*}
\WU_{ij}(t) = \UU_{ij}(t)-\Ent_{ij}(t)+\lceil t^{\alpha} \rceil .
\end{align*}
Then one can verify that $\WU_{ij}(t+1) = \WU_{ij}(t)+\WA_{ij}(t) - \RF_{ij}(t) $ and $\UU_{ij}(t)= \max\big\{ \WU_{ij}(t) -  \lceil t^{\alpha} \rceil , 0\big\}.$. Note that unlike $\UU_{ij}(t)$, $\WU_{ij}(t)$ can be negative.

\emph{On-demand Protocols with Virtual Requests:} At each time slot, the quantum switch attempts to create entanglement $\ket{\Psi_{ij}}$ using $\F_{ij}$ pairs of entanglement $\ket{\Psi_{0i}}$ and $\F_{ij}$ pairs of entanglement $\ket{\Psi_{0j}}$. The decisions $\{\F_{ij}\}_{i,j\in\Set{K}}$ need to satisfy the following constraints:
\begin{align}\nonumber
&\sum_{i\in\Set{K}} \F_{ij} \leq \Ent_{0j}(t)+\CC_{0j}(t), \quad j\in\Set{K} \\ \label{eq:on_demand_q}
&\F_{ij}  \leq  \Big\lceil \big(\WU_{ij}(t)+\WA_{ij}(t)\big)/q\Big\rceil, \quad i,j\in\Set{K} \\ \nonumber
&  \F_{ij}=\F_{ji} \in \mathbb{N},\quad i, j\in \Set{K} \\ 
&\Big(\Ent_{0i}(t) +\CC_{0i}(t)- \sum_{k\in\Set{K}} \F_{ik}\Big)\Big(\Ent_{0j}(t) +\CC_{0j}(t)- \sum_{k\in\Set{K}} \F_{kj}\Big)\nonumber\\
&\hspace{5cm} \cdot \bigg[\Big\lceil \big(\WU_{ij}(t)+\WA_{ij}(t)\big)/q \big)\Big\rceil- \F_{ij}\bigg] = 0,\quad i, j\in\Set{K}. \label{eq:equality_constraint_virtual}
\end{align}
Compared with the constraints \eqref{eq:entanglement_constraint} to \eqref{eq:equality_constraint}, there are two main differences. The first is the virtual requests in $\WA_{ij}(t)$, and the second is the factor $1/q$ in \eqref{eq:on_demand_q} and \eqref{eq:equality_constraint_virtual}. With this factor, the expected value $\RF_{ij}(t) $ is almost the same as $\big(\WU_{ij}(t)+\WA_{ij}(t)\big)/q$ provided that there are sufficient entanglement $\ket{\Psi_{0i}}$ and $\ket{\Psi_{0j}}$ for swapping. We can then show that zero average latency can be achieved by an on-demand protocol with virtual requests, denoted by $\widetilde{\pi}_{\mathrm{od}}$.

\subsection{Zero Average Latency of On-demand Protocols with Virtual Requests}

We can then show that the on-demand protocols with virtual requests achieve zero average latency whenever $(\lambda_{ij})$ is an interior point of $\V{\Lambda}$. The proof is organized in three steps.

Note that the goal is to prove $\mathbb{E}\big[\UU_{ij}(t)\big] \xrightarrow{ t \rightarrow \infty } 0$, which is equivalent to 
\begin{align*}
\lim_{t\rightarrow \infty}\mathbb{E}\big[\max\{\WU_{ij}(t+1)-\lceil (t+1)^\alpha\rceil, 0\}\big] = 0.
\end{align*}
For arbitrary random event $S_0(t)$ and $C_1(t)$, 
\begin{align}\nonumber
&\mathbb{E}\big[\max\{\WU_{ij}(t+1)-\lceil (t+1)^\alpha\rceil, 0\}\big] \\ \label{eq:new_three_term_1}
& =  \mathbb{E}\big[\max\{\WU_{ij}(t+1)-\lceil (t+1)^\alpha\rceil, 0\}| \overline{S_0}(t)\big] \mathbb{P}\big[\overline{S_0}(t)\big]\\  
&\quad+ \mathbb{E}\big[\max\{\WU_{ij}(t+1)-\lceil (t+1)^\alpha\rceil, 0\}|  {S_0}(t)\big] \mathbb{P}\big[ {S_0}(t) \big]. \label{eq:new_three_term_2}
\end{align}
We will then select $S_0(t)$, provide upper bounds for \eqref{eq:new_three_term_1} and \eqref{eq:new_three_term_2}, and show these upper bounds converge to 0 as $t$ goes to infinity.

\textbf{Step 1}: Select the random event $\overline{S_0}(t)$.

We consider a random event ${S}_0(t)$ defined as follows:
\begin{align}\label{eq:def_S_virtual}
{S}_0(t) = \Big\{ \sum_{i\in\Set{K}}\Big\lceil	\big[{\WU}_{ij}(t)+\WA_{ij}(t)\big]/q\Big\rceil\leq  {\Ent}_{0j}	(t)+\CC_{0j}(t)  , \forall j\in\Set{K}\Big\}
\end{align}
where ${\WU}_{ij}(t)$ and ${\Ent}_{0j}(t)$ are the number of backlog of entanglement request $\ket{\Psi_{ij}}$ and the number of entanglement $\ket{\Psi_{0j}}$ achieved by the used on-demand protocol. If the event $S_0(t)$ occurs, then the quantum switch has sufficient entanglement to address the entanglement request, i.e., $\F_{ij}(t) \geq \Big\lceil\big[ \WU_{ij}(t)+\WA_{ij}(t)\big]/q\Big\rceil$, $i,j\in\Set{K}$ if an on-demand protocol with virtual requests is used.

\textbf{Step 2}: Provide an upper bound for \eqref{eq:new_three_term_1}.

Note that
\begin{align*}
 \mathbb{E}\big[\max\{\WU_{ij}(t+1)-\lceil (t+1)^\alpha\rceil, 0\}| \overline{S_0}(t)\big] \mathbb{P}\big[\overline{S_0}(t)\big] 
 \leq \mathbb{E}\Big[ \sum_{\tau = 0}^{t}\WA_{ij}(\tau)|\overline{S_0}(t)\Big]\mathbb{P}\big[\overline{S_0}(t)\big]
 \end{align*}
which converges to 0 because of the following lemma.
\begin{lemma}\label{lemma:lemma_long_virtual}
Under Assumption 2, if an on-demand protocol is used,
then for any $\tilde{i}$, $\tilde{j}\in\Set{K}$
\begin{align}\label{eq:conditional_exp_A0_virtual}
\lim_{t\rightarrow \infty}\mathbb{E}\Big[ \sum_{\tau = 0}^{t}\WA_{\tilde{i}\tilde{j}}(\tau)|\overline{S_0}(t)\Big]\mathbb{P}\big[\overline{S_0}(t)\big]=0.
\end{align}
\end{lemma}
The proof of Lemma \ref{lemma:lemma_long_virtual} can be found in Appendix \ref{apd:lemma_long_virtual}. 

\textbf{Step 3}: Provide an upper bound for \eqref{eq:new_three_term_2}.

To obtain the  upper bound, we need to introduce several random events:
\begin{align*}
&C_1(t):\exists i_0, j_0,~ \frac{1}{t+1}\sum_{\tau = 0}^{t}\WA_{i_0j_0}(\tau) >\lambda_{i_0j_0} +\tilde{\epsilon}.
\end{align*}
 
 Then we can rewrite \eqref{eq:new_three_term_2}:
 \begin{align} 
 \nonumber
  &\mathbb{E}\big[\max\{\WU_{ij}(t+1)-\lceil (t+1)^\alpha\rceil, 0\}|  {S_0}(t)\big] \mathbb{P}\big[ {S_0}(t) \big] \\ \nonumber
 &= \mathbb{E}\big[\max\{\WU_{ij}(t+1)-\lceil (t+1)^\alpha\rceil, 0\}|  {S_0}(t)\cap {C}_1(t)\big] \mathbb{P}\big[ {S_0}(t)\cap  {C}_1(t)\big] \\
 &\quad +\mathbb{E}\big[\max\{\WU_{ij}(t+1)-\lceil (t+1)^\alpha\rceil, 0\}|  {S_0}(t)\cap \overline{C}_1(t) \big] \mathbb{P}\big[ {S_0}(t)\cap \overline{C}_1(t). \big]\label{eq:new_three_term}
 \end{align}
 One can verify that all the terms above converge to 0 as $t$ goes to infinity, and the proof can be found in Appendix \ref{apd:new_three_term}.  This gives the following theorem. 
 
\begin{theorem}\label{thm:on_demand_general_virtual}
The quantum switch system achieves zero average latency using an on-demand protocol with virtual requests $\widetilde{\pi}_{\mathrm{od}}$ if $(\lambda_{ij})_{i,j\in\Set{K}}$ is an interior point in $\V{\Lambda}$ under Assumption 2.
\end{theorem}

\section{Experimental Results}\label{sec:numerical_results}
In this section, we investigate the performance of the proposed protocols, namely, the stationary, the max weight, and the on-demand protocols with a quantum network discrete event simulator, NetSquid \cite{Cooetal:20}. Since the stability of these protocols are proven in previous sections, we focus on other performance metrics in practical scenarios. Moreover, we relax several assumptions including the infinite amount of memory slots for storing \ac{epr} pairs and the infinite qubit lifetime.


\subsection{Simulation Setting}
The quantum switch can be implemented with different technologies. For example, qubits stored in quantum memories can be realized with electron spins of SiV defect centers \cite{Bhaetal:20}. The quantum switch is equipped with photon sources, and each of them generates a pair of entangled photons in each time slot. One of the photons interacts with the electron spin, and the other photon is sent to one of the end nodes through a quantum channel and interacts with the electron spin at the end node. The photon-spin interaction is essentially a CNOT operation on the photon and the spin. After the photon-spin interaction, photons are measured in the X basis by beamsplitters and  photon detectors. The entanglement swapping operation in the quantum switch can also be realized with photon-spin interactions. To be consistent with practical devices, we assume that qubits stored in the memory suffer from decoherence. Furthermore, we assume that the quantum switch has a finite number of memory slots, and these memory slots are equally distributed among interfaces. With these practical constraints, we can evaluate more performance metrics in addition to the stability or throughput of the quantum switch system. In particular, we are interested in the following figures of merit:
\begin{itemize}
\item Average fidelity: for a state ${\rho}$ shared between node $i$ and $j$, the fidelity is defined as $F(\rho) =\bra{\Psi_{ij}} \rho\ket{\Psi_{ij}}$. Note that a generated \ac{epr} pair may decohere in the memory slots before being used to address entanglement requests. 
\item Average latency: the latency of an entanglement request is defined as the amount of time to address the entanglement request. Note that the entanglement request may occur at any time in a slot, and we evaluate the latency in units of nanoseconds rather than slots. 
\end{itemize}

The protocols developed in previous sections do not specify how to prioritize the  \ac{epr} pairs for entanglement swapping and entanglement requests  to address. The prioritization of  the  \ac{epr} pairs and which entanglement requests can impact fidelity and latency described above. In this section, we apply the \ac{fifo} method to process  entanglement requests, i.e., the oldest entanglement request is the first to address. Regarding the order of \ac{epr} pairs, we consider two methods: \ac{oqf} and \ac{yqf}, which use the oldest and youngest qubit for entanglement swapping, respectively. All the protocols discard \ac{epr} pairs when their fidelities fall below a preset threshold. This is equivalent to discarding \ac{epr} pairs after a fixed amount of time.

Each time slot is 1000 ns long. The number of entanglement requests between any two end nodes $i$ and $j$ in a time slot is given by a mixture of two Poisson distributions with different rates. Specifically, $\A_{ij}(t)$ is i.i.d over time and is given by $\A_{ij}(t) = Z \cdot Y_{1} + (1-Z) \cdot Y_{2}$, where $Z$ is a Bernoulli random varable with mean $1/2$, and $Y_1$ and $Y_2$ are independent Poisson random variables with mean $\lambda_1$ and $\lambda_2$. The rate of this mixed Poisson process is then $\lambda_{ij} = (\lambda_{1}+\lambda_{2})/2$. Throughout this section, the rate of entanglement requests is the same for all end node pairs.  
 

\subsection{Performance Analysis}
In this subsection, unless otherwise specified, the number of memory slots is 100 per interface and the entanglement swapping probability $q=0.9$. The channels between the switch and end nodes are lossy optical fibers, and the entanglement generation probability $p=0.9$ is the same for all interfaces. Note that $p=0.9$ corresponds to the distance between the switch and end nodes being 2.3 km given that the fiber attenuation coefficient is 0.2 dB/km.\footnote{When $p=0.9$ (i.e., the distance between the switch and end nodes is 2.3 km), the link level entanglement rate is 0.9 \ac{epr} pair per $\mu$s. This is reasonable since ideally photon sources can emit photons at the rate of $10^7$ Hz, corresponding to 10 \ac{epr} pairs per $\mu$s.} The qubits suffer from dephasing noise \cite{Pre:B} when staying idle in memory slots, and the T2 time for the dephasing noise in each memory slot is set to 1 millisecond. Specifically, the dephasing noise model in a memory slot is modelled as follows:
\begin{align*}
   \mathcal{N}_{\mathrm{dephase}}:\quad \rho \rightarrow (1-p_\mathrm{dephase})\rho +p_\mathrm{dephase}\sigma_Z \rho \sigma_Z
\end{align*}
where $\rho$ is the density matrix of a qubit, $\sigma_Z = \ket{0}\bra{0} - \ket{1}\bra{1}$ is one of the Pauli operators, and $p_{\mathrm{dephase}}$ is the dephasing probability, given by $  p_{\mathrm{dephase}} = (1-\exp{-\Delta t/T_2})/2$, in which $\Delta t$ denotes the time that a qubit stays idle in the memory slot. If one qubit of an \ac{epr} $\ket{\Psi}$ is stored in a memory qubit, then after time $\Delta t$, one can verify that its fidelity becomes
\begin{align*}
    \bra{\Psi} \big[(1-p_\mathrm{dephase})\ket{\Psi}\bra{\Psi} +p_\mathrm{dephase}(\sigma_Z\otimes I ) \ket{\Psi}\bra{\Psi} (\sigma_Z\otimes I )\big] \ket{\Psi} = \frac{1}{2}(1+\exp{-\Delta t/T_2}).
\end{align*}

\renewcommand{\arraystretch}{1.5}
\begin{table}[htbp]
\vspace{-0mm}
\centering
\begin{tabular}{  p{4.5cm}  | p{4.2cm}  | p{4.2cm}}
\hline
\textbf{Protocol}& \textbf{Average Fidelity}& \textbf{Average Latency ($\mu$s)}\\
\hline
\rowcolor{blue!10!white} Stationary (\ac{yqf}) & 0.976 & 15.6\\
Stationary (\ac{oqf}) & 0.908 & 14.2\\
\rowcolor{blue!10!white} On-demand (\ac{yqf}) &  0.975  &  12.4  \\
On-demand (\ac{oqf}) & 0.916  & 14.4 \\
\hline
\end{tabular}
\vspace{3mm}\caption{Performance of Entanglement Swapping Protocols: $\lambda_{ij} = 0.2/\mu\mathrm{s}, \forall i, j\in\Set{K}$.}
\label{tab:results_heavy}
\end{table}

\renewcommand{\arraystretch}{1.5}
\begin{table}[htbp]
\vspace{-0mm}
\centering
\begin{tabular}{  p{4.5cm}  | p{4.2cm}  | p{4.2cm}}
\hline
\textbf{Protocol}& \textbf{Average Fidelity}& \textbf{Average Latency ($\mu$s)}\\
\hline
\rowcolor{blue!10!white} Stationary (\ac{yqf}) & 0.961 & 0.092\\
Stationary (\ac{oqf}) & 0.752 & 0.089\\
\rowcolor{blue!10!white} On-demand (\ac{yqf}) &  0.960  &  0.080  \\
On-demand (\ac{oqf}) & 0.752  & 0.067 \\
\hline
\end{tabular}
\vspace{3mm}\caption{Performance of Entanglement Swapping Protocols: $\lambda_{ij} = 0.12/\mu\mathrm{s}, \forall i, j\in\Set{K}$.}
\label{tab:results_light}
\end{table}

We begin with a comparison of the stationary protocol and an on-demand protocol. For each protocol, we further implement the \ac{yqf} and \ac{oqf} methods to prioritize \ac{epr} pairs. The number of interfaces is $K=5$. Tables \ref{tab:results_heavy} and \ref{tab:results_light} show the performance of the developed protocols for $\lambda_{ij} = 0.2/\mu\mathrm{s}$ and $\lambda_{ij} = 0.12/\mu\mathrm{s}$, $\forall i, j\in\Set{K}$, respectively. First, the on-demand protocols and stationary protocols perform similarly in terms of fidelity. Regarding latency, the on-demand protocols perform the best in most cases. The only exception is 
when $\lambda_{ij} = 0.2/\mu\mathrm{s}$, where the stationary (OQF) performs slightly better than the on-demand (OQF). Note that the on-demand protocol does not require any statistical knowledge of the arrival process or the systems, and that it involves low computational overhead. Together with their desirable performance, on-demand protocols are desirable choices in practice. Second, the average fidelities of the \ac{yqf} protocols are much higher than those of the \ac{oqf} protocols, but the average latencies are generally greater than those of the \ac{oqf} protocols, especially when $\lambda_{ij}$ is small. This agrees with intuition. \ac{yqf} tends to use newly generated \ac{epr} pairs for entanglement swapping, so the fidelities are higher. Moreover, since the \ac{epr} pairs are discarded when their fidelities are low, more \ac{epr} pairs are discarded when \ac{yqf} is used. Since there are fewer \ac{epr} pairs to use, latencies increase. We next use the on-demand protocol  as the default protocol to evaluate the performance of quantum switch in different settings. 

 \begin{figure}[t]
\center	
\includegraphics[width=0.7\linewidth, draft=false]{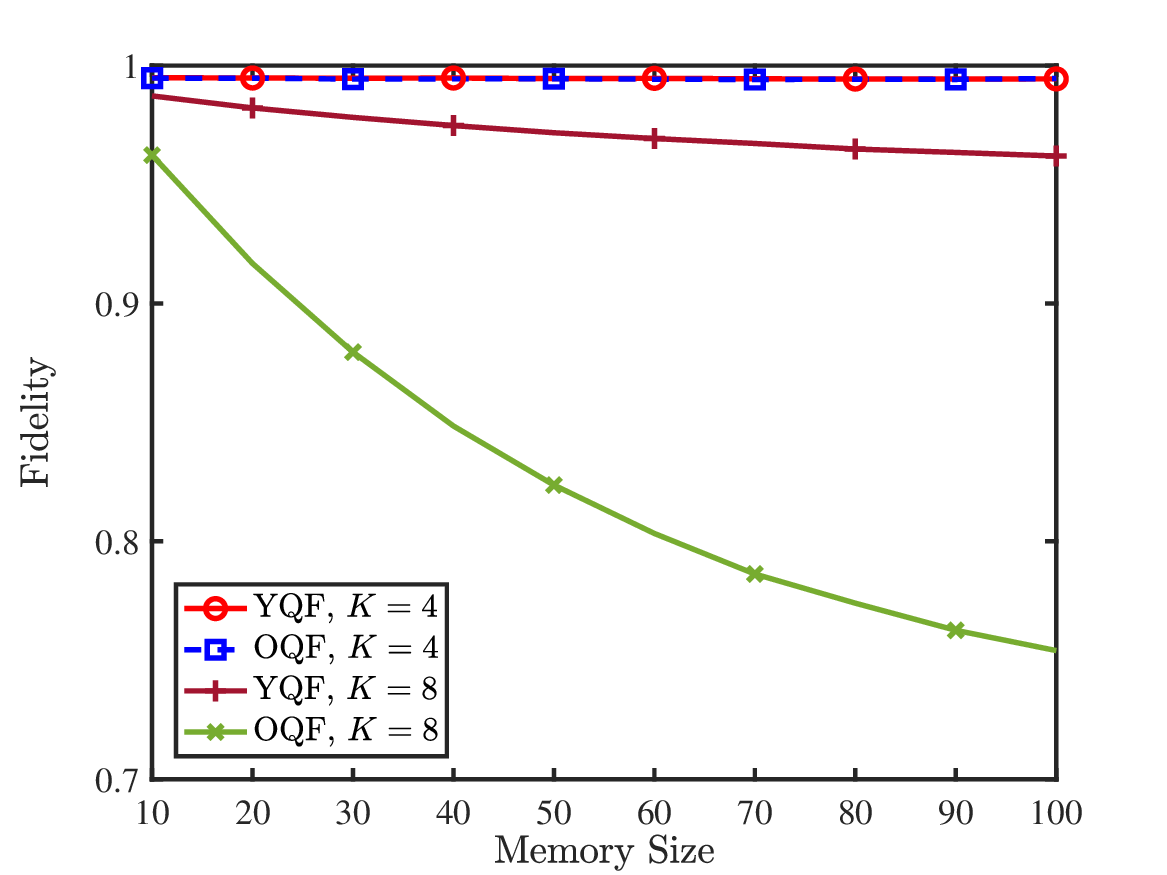}
\caption{Average fidelity as a function of the memory size. The x-axis denote the number of memory slots normalized by $K$. }
\label{fig:fidelity_memorysize}
\end{figure}

 \begin{figure}[t]
\center	
\includegraphics[width=0.7\linewidth, draft=false]{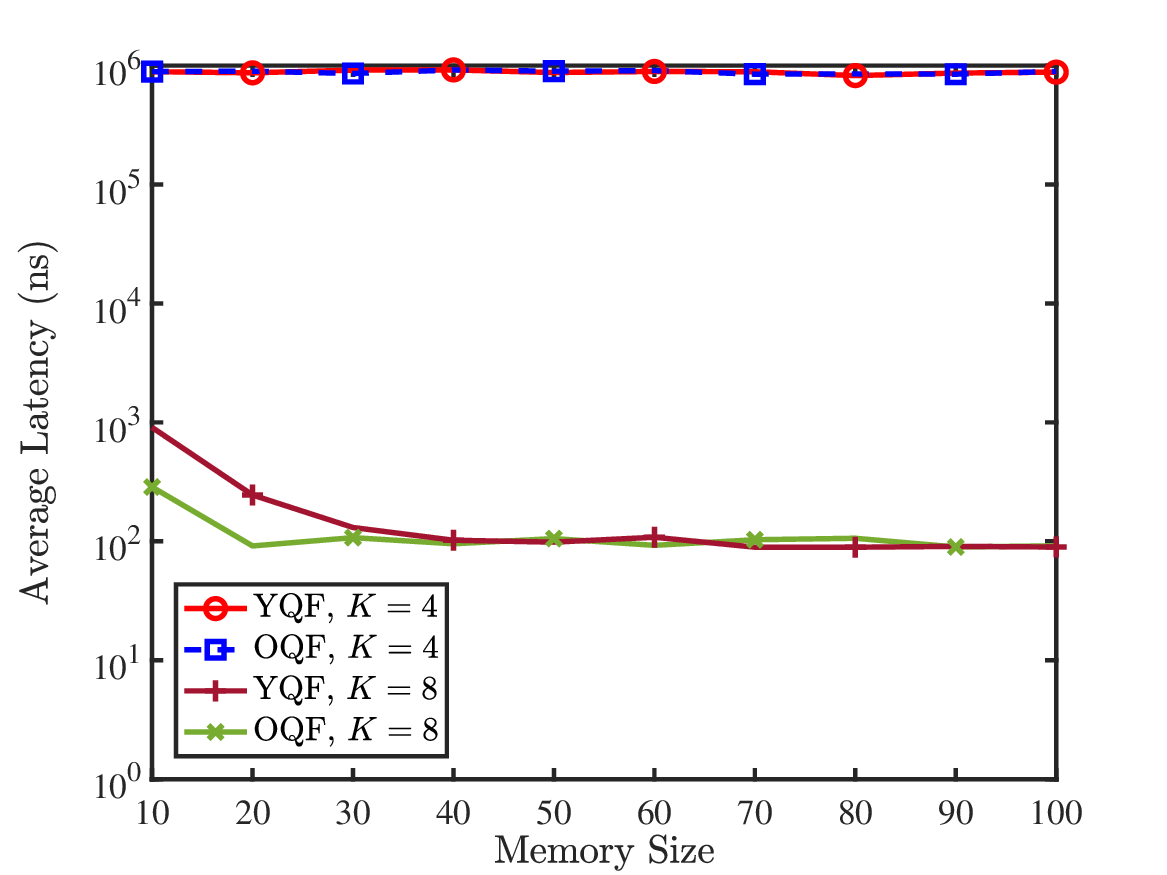}
\caption{Average latency as a function of the memory size. The x-axis denote the number of memory slots normalized by $K$. }
\label{fig:latency_memorysize}
\end{figure}

Figures \ref{fig:fidelity_memorysize} and \ref{fig:latency_memorysize} respectively show  average fidelity and latency of the quantum switch achieved by the on-demand protocols (\ac{yqf} and \ac{oqf}) as functions of the number of memory slots. In these two figures, we set $\sum_{i,j}\lambda_{ij}=1.2/\mu\mathrm{s}$. First, for $K=8$, the average fidelity and average latency decrease with memory size; for $K=4$, the average fidelity and average latency remain constant. This is because for $K=4$, the rates $(\lambda_{ij})_{1\leq i, j \leq 4}$ lie outside the capacity region. Therefore, the quantum switch is not stable, leading to high latencies. Since there are many unaddressed entanglement requests most of the time, the generated qubits between the switch and end nodes are used immediately for entanglement swapping, and this leads to a fidelity close to one. An increase in memory slots results in qubits staying in the memory longer before consumed, leading to a decrease in fidelity. In addition, more memory slots imply that the quantum switch discards fewer \ac{epr} pairs and therefore reduces latency. Second, the latency reduction due to the increase in memory size becomes less significant when the number of the memory slots goes beyond 30. This is because, when the number of the memory slots is sufficiently large, the quantum switch can accommodate enough \ac{epr} pairs to address the entanglement requests, and more memory slots provide diminishing marginal benefit. 


\begin{figure}[t]
\center	
\includegraphics[width=0.7\linewidth, draft=false]{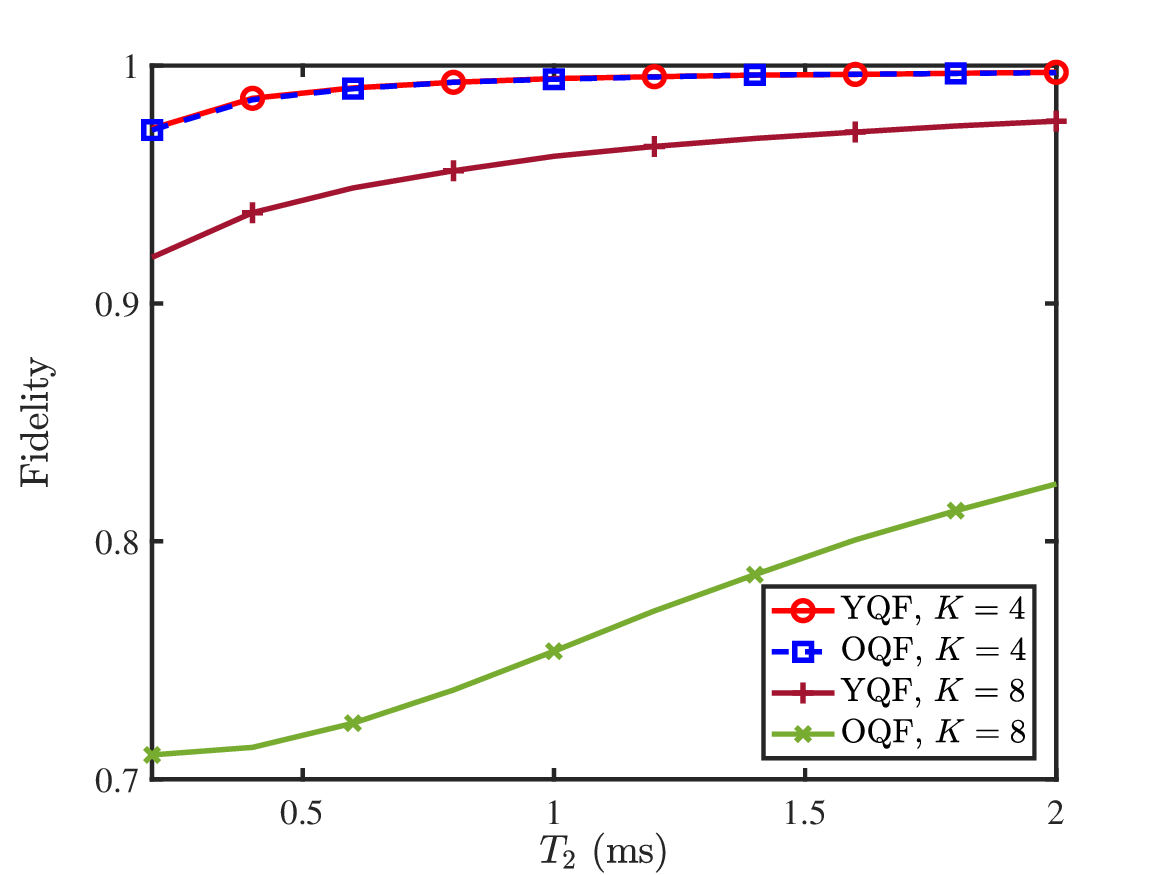}
\caption{Average fidelity as a function of $T_2$. }
\label{fig:fidelity_T2}
\end{figure}

 \begin{figure}[t]
\center	
\includegraphics[width=0.7\linewidth, draft=false]{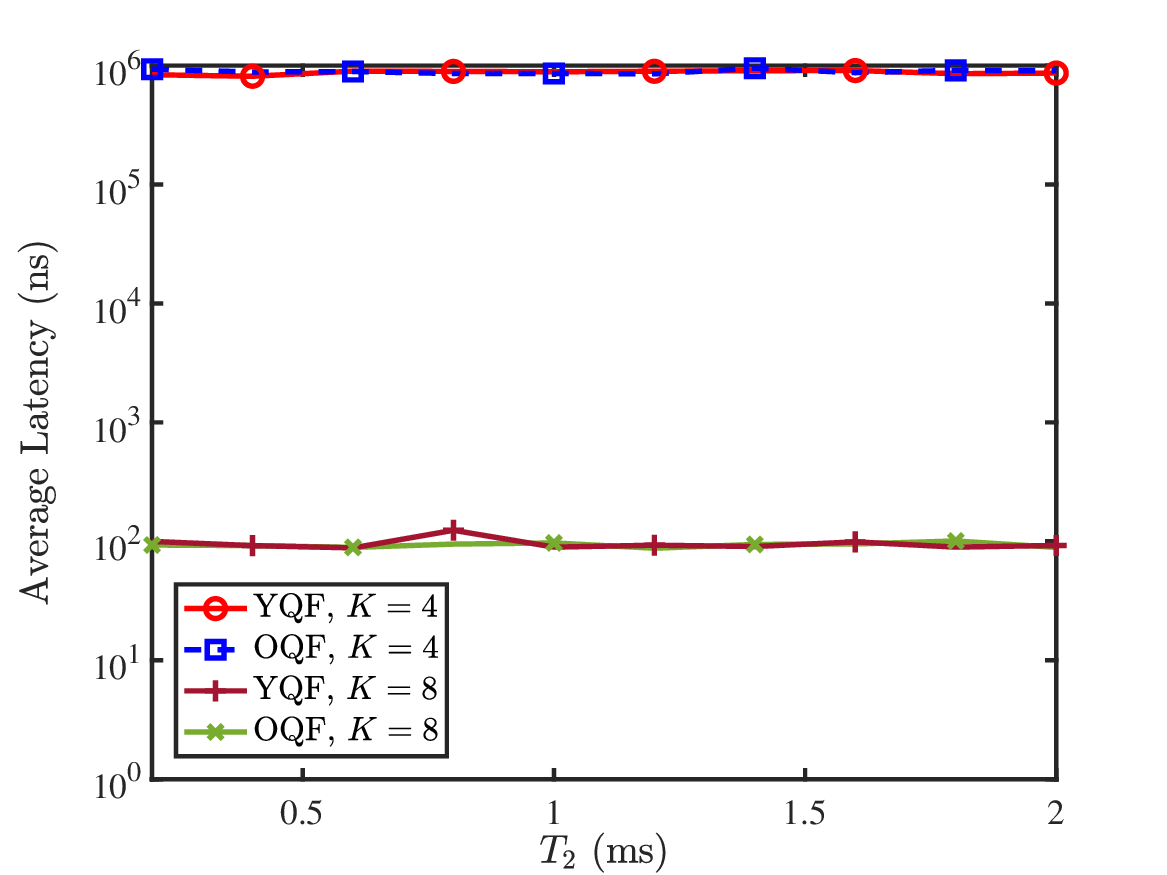}
\caption{Average latency as a function of $T_2$. }
\label{fig:latency_T2}
\end{figure}

Figures \ref{fig:fidelity_T2} and \ref{fig:latency_T2} respectively show average fidelity and latency of the quantum switch achieved by the on-demand protocols (\ac{yqf} and \ac{oqf}) as functions of decoherence time $T_2$. In these two figures, we set $\sum_{i,j}\lambda_{ij}=1.2/\mu\mathrm{s}$. First, fidelity increases with $T_2$. This is consistent with the dephasing noise model shown earlier, where we can find that the dephasing probability decreases with $T_2$. Second, latency remains constant as a function of $T_2$. In the considered setup, $T_2$ has little effect on the amount or the timing of entanglement generation, and thus the latencies of entanglement requests hardly depend on $T_2$. Third, similarly to Figures \ref{fig:fidelity_memorysize} and \ref{fig:latency_memorysize},  \ac{yqf} and \ac{oqf} have almost the same performance when $K=4$. Again this is because the quantum switch is not stable for $K=4$, and thus the generated qubits between the switch and end nodes are used immediately so that the \ac{yqf} and \ac{oqf} give the same choice of \ac{epr} pairs.

 \begin{figure}[t]
\center	
\includegraphics[width=0.7\linewidth, draft=false]{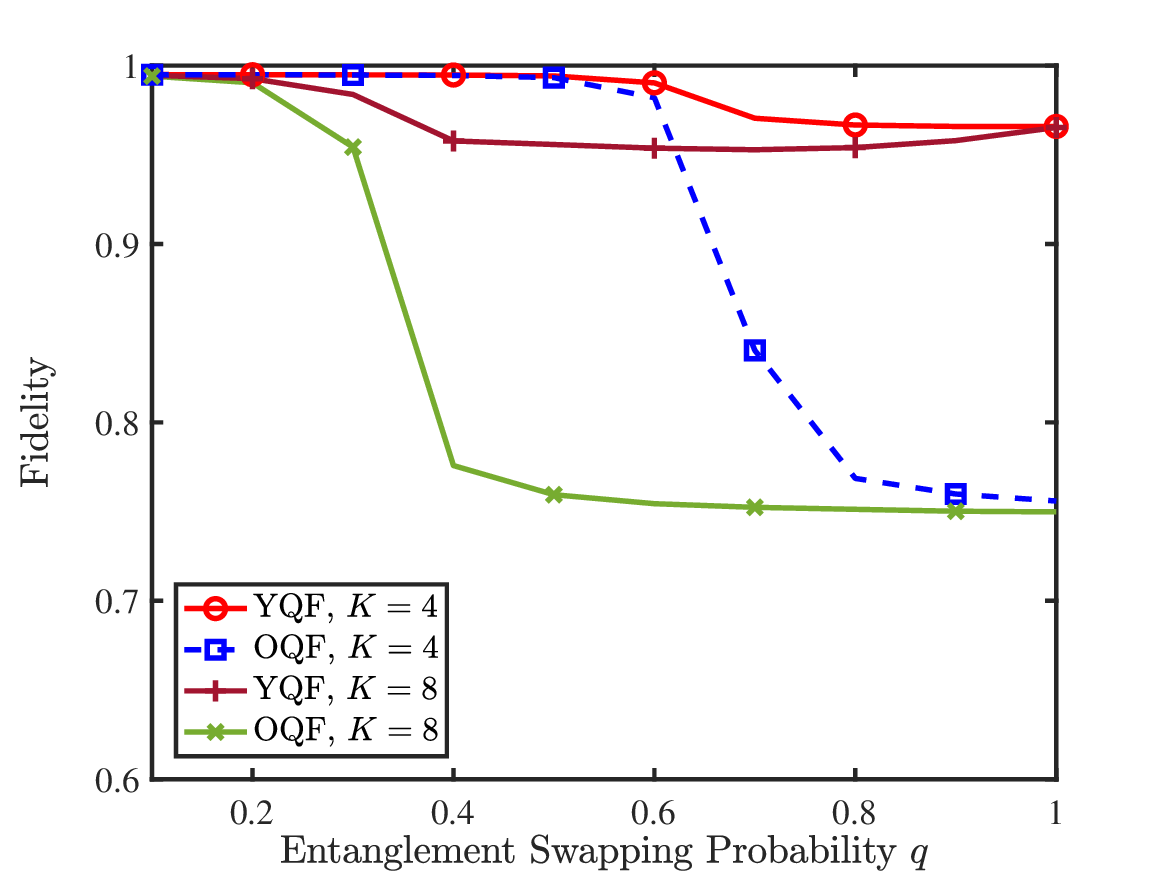}
\caption{Average fidelity as a function of entanglement swapping success probability $q$.}
\label{fig:fidelity_q}
\end{figure}

 \begin{figure}[t]
\center	
\includegraphics[width=0.7\linewidth, draft=false]{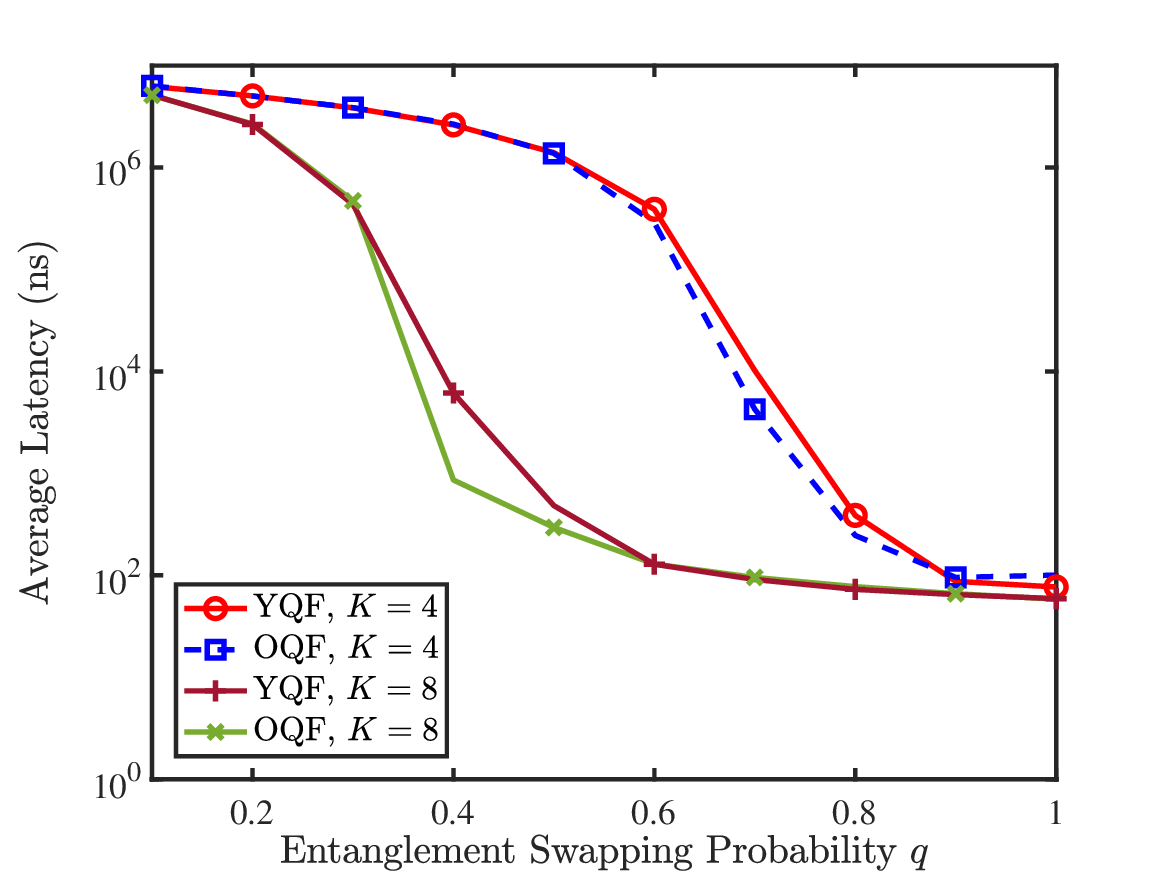}
\caption{Average latency as a function of entanglement swapping success probability $q$.}
\label{fig:latency_q}
\end{figure}

Figures \ref{fig:fidelity_q} and \ref{fig:latency_q} respectively show average fidelity and latency of the quantum switch achieved by the on-demand protocols (\ac{yqf} and \ac{oqf}) as functions of the entanglement swapping success probability $q$. In these two figures, we set $\sum_{i,j}\lambda_{ij}=2/\mu\mathrm{s}$. First, fidelity initially decreases with $q$ and then remains constant. This is because for small $q$, the rates are outside the capacity region, and the generated qubits between the switch and end nodes are used immediately for entanglement swapping, leading to fidelities close to one. As $q$ increases, the rates gradually move into the capacity region, and some qubits stay in the memory before being consumed for entanglement swapping, leading to a decrease in fidelity. When $q$ is sufficiently large, the memory slots are full most of the time, and increasing $q$ does not significantly improve the memory occupancy rate. In these occasions, average fidelity does not change with $q$. Second, average latency first deceases with $q$ and then remains constant. The reasoning for this behavior is almost the same as that for fidelity. Third, one observes that the sudden change of fidelity and latency occurs around $q=0.33$ for $K=8$ and $q=0.67$ for $K=4$. These two points exactly correspond to the boundary points for the capacity region. This shows that the switch demonstrate entirely different behavior inside or outside the capacity region, which is consistent with the stability analysis in earlier sections. This figure also shows that if $q$ corresponds to a boundary point of the capacity region, increasing $q$ provides limited performance improvement.

\section{Conclusion}\label{sec:conclude}
We develop efficient entanglement swapping protocols for a quantum switch and analyze their performance in terms of stability of the switch, fidelity of \ac{epr} pairs and latency of entanglement requests. We determine the capacity region for entanglement rates under the assumption of infinite memory size and infinite qubit lifetime. Specifically, we show that no entanglement swapping protocols can stabilize the switch if the entanglement rates lie outside this capacity region, and develop different protocols that stabilize the switch when the entanglement rates correspond to an interior point of the capacity region. The performance of the developed protocols is evaluated using NetSquid, and practical constraints such as decoherence in memory slots are accounted for. 
We show that stationary protocols and on-demand protocols exhibit high fidelity and low latency. Moreover, the tradeoff between memory size, decoherence rate, fidelity, and latency shown in the simulation results offer guidance in the implementation of quantum switches.

A potential future direction is the design and analysis of switches with a finite memory size and finite lifetime. Note that in this manuscript, simulation results are obtained in practical scenarios, whereas the stability analysis is based on the assumptions that the switch has sufficiently many memory slots and that qubits have infinite lifetime when stored in the memory. The capacity region for a finite memory size and lifetime must be different, and the entanglement swapping protocols must be designed accordingly. 


%

\section*{Acknowledgment}
The authors wish to thank Philippe Nain for his helpful suggestions and careful reading of the manuscript.

\bibliographystyle{IEEEtran}
\bibliography{Bibliography/IEEEabrv,Bibliography/StringDefinitions,Bibliography/BiblioCV,Bibliography/WINS-Quantum,Bibliography/Wgroup,temp}
\appendices
\section{Proof of Theorem \ref{thm:capacity_region}}\label{apd:proof_capacity_region}

Here we prove that $(\lambda_{ij})\in\V{\Lambda}$ is necessary for the quantum switch to be stable. The following lemma is inspired by Lemma 1 in \cite{NeeThe:03p}.
\begin{lemma}\label{lemma:queue_necessary_stability}
If a quantum switch is stable, then for any $\delta>0$, there exits a finite value $V$ such that for any $t_0>0$, there exists $\tilde{t}>t_0$ such that
\begin{align}\label{eq:tilde_t_infinite}
\mathbb{P}\big[ \UU_{ij}(\tilde{t}) \leq V\big] > 1-\delta, \quad \forall i,j\in\Set{K}.
\end{align}
In particular, for the case $\delta = 1/2$, there exists a value $V$ such that the probability that the unfinished requests in all queues, i.e., $\UU_{ij}(t)$, $i,j\in\Set{K}$, simultaneously drop below $V$ is greater than $1/2$ infinitely often.
\end{lemma}

Consider a quantum switch with request rate matrix $(\lambda_{ij})_{i,j\in \Set{K}}$. Suppose there exists an entanglement swapping algorithm that stabilizes the quantum switch. We need to show that there exist variables $\{f_{ij}\}_{i,j\in \Set{K}}$ satisfying \eqref{eq:capacity_necessary_1}-\eqref{eq:capacity_necessary_3}. With the notations defined in Section \ref{sec:system_dynamics}, we have for arbitrary time slot $t \ge 0$,
\begin{align}
\sum_{\tau =0}^{t-1}\sum_{i\in \Set{K}} \F_{ij}(\tau)& \leq  \sum_{\tau =0}^{t-1} \CC_{0j}(\tau), \quad \forall j\in \Set{K} \label{eq:constraint_C} \\
\sum_{i,j\in \Set{K}}F_{ij}(t) &\leq \BSN \label{eq:constraint_L} \\
\UU_{ij}(t) &=  \max\bigg\{\sum_{\tau = 0}^{t-1}\A_{ij}(\tau)  - \sum_{\tau = 0}^{t-1} \RF_{ij}(\tau) ,0\bigg\},  \quad \forall i,j\in \Set{K}.\label{eq:constraint_U}
\end{align}

Fix arbitrarily small $\epsilon>0$, and set $\delta = 1/2$. According to Lemma \ref{lemma:queue_necessary_stability}, there exists finite $V$ and $\tilde{t}$ such that $\tilde{t}\geq V/\epsilon$ and
\begin{align}
\mathbb{P}\big[\UU_{ij}(\tilde{t})\leq V\big] > 1/2. 
\end{align}
Hence,
\begin{align*}
\mathbb{P}\big[{\UU_{ij}(\tilde{t})}/{\tilde{t}}\leq \epsilon\big] &= \mathbb{P}\big[\UU_{ij}(\tilde{t})\leq \tilde{t}\epsilon\big]\geq \mathbb{P}\big[\UU_{ij}(\tilde{t})\leq V\big] \geq 1/2.
\end{align*}
We then define the following event $A_0$: ${\UU_{ij}(\tilde{t})}/{\tilde{t}}\leq \epsilon, \quad \forall i,j\in \Set{K}$.
It follows that $\mathbb{P}[A_0]\geq 1/2$.


We now consider the following events:
\begin{align}\label{eq:event_A}
&A_1:~ \frac{1}{\tilde{t}}\sum_{\tau=0}^{\tilde{t}-1} \A_{ij}(\tau) \geq \lambda_{ij}-\epsilon, \quad \forall i,j\in \Set{K} \\ \label{eq:event_C}
&A_2:~\frac{1}{\tilde{t}}\sum_{\tau=0}^{\tilde{t}-1} \CC_{0i}(\tau)  \leq p_i + \epsilon, \quad \forall  i\in\Set{K} \\ \label{eq:event_q}
&A_3:~\frac{ \sum_{\tau=0}^{\tilde{t}-1} \RF_{ij}(\tau) }{  \sum_{\tau=0}^{\tilde{t}-1} \F_{ij}(\tau) 	} \leq q + \epsilon, \quad \forall i,j\in \Set{K}.
\end{align}
Since $[\sum_{t = 0}^{t-1}\A_{ij}(t)]/t$ converges to $\lambda_{ij}$ almost surely due to Birkhoff-Khinchin theorem, it converges to $\lambda_{ij}$ in probability. This means that $\mathbb{P}[A_1]\geq 0.99$ for sufficiently large $\tilde{t}$. Similarly, since $\{{\CC}_{0i}(t):t\geq 0\}$ are i.i.d with respect to $t$, $\mathbb{P}[A_2]\geq 0.99$ for sufficiently large $\tilde{t}$. Regarding the event $A_3$, since entanglement swapping succeeds with probability $q$, the probability of $A_3$  is no less than 0.99 for sufficiently large  $  \sum_{\tau=0}^{\tilde{t}-1} \F_{ij}(\tau) $. Note that when events $A_0$ and $A_1$ occur simultaneously, $\forall i,j \in \Set{K}$
\begin{align*}
 \sum_{\tau=0}^{\tilde{t}-1} \F_{ij}(\tau) &\geq  \sum_{\tau=0}^{\tilde{t}-1} \RF_{ij}(\tau) \geq  \bigg[\sum_{\tau=0}^{\tilde{t}-1} \A_{ij}(\tau) \bigg]-  \UU_{ij}(\tilde{t}) \geq ( \lambda_{ij}-\epsilon)\tilde{t} - V
\end{align*}
where the second inequality comes from \eqref{eq:constraint_U}, and the last inequality follows from the joint occurrence of the events $A_0$ and $A_1$. Hence, $  \sum_{\tau=0}^{\tilde{t}-1} \F_{ij}(\tau)$ can be made arbitrarily large through choice of $\tilde{t}$. Consider the joint event $\cap_{l=0}^3 A_l$; its probability can be lower bounded as follows:
\begin{align}\nonumber
\mathbb{P}\big[\cap_{l=0}^3 A_l\big] &= 1 - \mathbb{P}\big[\cup_{l=0}^3 \overline{A}_l\big]  \geq \frac{1}{2} -  \mathbb{P}\big[\cup_{l=1}^3 \overline{A}_l\big]  \geq \frac{1}{2} -  \mathbb{P}\big[\overline{A}_1\big] -   \mathbb{P}\big[\overline{A}_2\big] - \mathbb{P}\big[{A}_1\cap {A}_2\cap \overline{A}_3\big] \\
&\geq \frac{1}{2} - 0.01 - 0.01 - 0.01>0. \label{eq:cap_nonzero}
\end{align}

Now define
\begin{align*}
f_{ij}(\tilde{t}) = \frac{1}{\tilde{t}} \sum_{\tau = 0}^{\tilde{t}-1} \F_{ij}(\tau) \quad \forall i, j\in \Set{K}
\end{align*}
and the following events:
\begin{align}
&A_4:~\sum_{i\in \Set{K}}f_{ij}(\tilde{t})  \leq p_j +\epsilon , \quad \forall j\in \Set{K} \label{eq:flow_constraint_C} \\
&A_5:~\sum_{i,j\in \Set{K}}f_{ij}(\tilde{t})  \leq \BSN \label{eq:flow_constraint_L} \\
&A_6:~\lambda_{ij}-\epsilon \leq \epsilon + (q+\epsilon) f_{ij}(\tilde{t})  , \quad \forall i, j\in\Set{K}\label{eq:flow_constraint_q}.
\end{align}
We next show that $\cap_{l=0}^3 A_l \subseteq  A_4\cap A_5 \cap A_6$. In fact, if $\cap_{l=0}^3 A_l $ occurs, we will show that $A_4\cap A_5 \cap A_6$ occurs. For all $j$,
\begin{align*}
\sum_{i\in \Set{K}}f_{ij}(\tilde{t})    =   \frac{1}{\tilde{t}}\sum_{\tau = 0}^{\tilde{t} - 1}\sum_{i\in \Set{K}}\F_{ij}(\tau)\leq  \frac{1}{\tilde{t}}\sum_{\tau = 0}^{\tilde{t} - 1}\CC_{0j}(\tau) \leq p_j +\epsilon 
\end{align*}
where the first inequality is because of \eqref{eq:constraint_C} and the second inequality is because $A_2$ occurs. Hence, event $A_4$  occurs. Moreover,
\begin{align*}
\sum_{i,j\in \Set{K}}f_{ij}(\tilde{t})  =    \frac{1}{\tilde{t}}\sum_{\tau = 0}^{\tilde{t} - 1}\sum_{i\in \Set{K}}\F_{ij}(\tau) \leq \BSN
\end{align*}
where the inequality comes from \eqref{eq:constraint_L}. Hence, event $A_5$ occurs. In addition, for all $i,j\in \Set{K}$
\begin{align*}
\lambda_{ij}-\epsilon & \leq \frac{1}{\tilde{t}} \sum_{\tau = 0}^{\tilde{t}-1}\A_{ij}(\tau) \leq \frac{1}{\tilde{t}}\bigg[ \UU_{ij}(\tilde{t}) + \sum_{\tau = 0}^{\tilde{t}-1} \RF_{ij}(\tau) \bigg] \leq  \epsilon + \frac{q+\epsilon}{\tilde{t}} \sum_{\tau = 0}^{\tilde{t}-1}\F_{ij}(\tau) \leq  \epsilon + (q+\epsilon) f_{ij}(\tilde{t})
\end{align*}
where the first inequality is because $A_1$ occurs, the second inequality comes from \eqref{eq:constraint_U}, and the third inequality is because $A_0$ and $A_3$ occur. Hence, the event $A_6$ occurs. This shows that $\cap_{l=0}^3 A_l\subseteq  A_4\cap A_5 \cap A_6 $. Therefore,
$
\mathbb{P}[A_4\cap A_5\cap A_6] \geq \mathbb{P}\big[\cap_{l=0}^3 A_l \big] >0
$, where the last inequality is because of \eqref{eq:cap_nonzero}. Then, with nonzero probability, the variables $\{f_{ij}(\tilde{t})\}_{i,j\in \Set{K}}$ defined based on $\F_{ij}(\tau)$ satisfy \eqref{eq:flow_constraint_C}-\eqref{eq:flow_constraint_q}. It follows that for arbitrarily small $\epsilon>0$, there exist $\{f_{ij}\}_{i,j\in \Set{K}}$  that satisfy 
\begin{align*}
&\sum_{i\in \Set{K}}f_{ij}  \leq p_j +\epsilon , \quad \forall j\in \Set{K}  \\
&\sum_{i,j\in \Set{K}}f_{ij}  \leq \BSN  \\
&\lambda_{ij}-\epsilon \leq \epsilon + (q+\epsilon) f_{ij} , \quad \forall i, j\in\Set{K}.
\end{align*}
This proves that $(\lambda_{ij})_{i,j\in\Set{K}}$ is a limit point in the capacity region $\V{\Lambda}$ defined in \eqref{eq:capacity_necessary_1}-\eqref{eq:capacity_necessary_3}. Since $\V{\Lambda}$ is compact, it follows that $(\lambda_{ij})_{i,j\in\Set{K}}\in \V{\Lambda}$.

\section{Proof of Lemma \ref{lemma:T_step_Lyapunov_stable}}\label{apd:T_step_Lyapunov_stable}
Without loss of generality, we assume $J=0$. Consider \eqref{eq:condition_T_LD} at times $t=mT$. Taking expectations of \eqref{eq:condition_T_LD} over the distribution of $\M{U}(mT)$ and summing over $m$ from $m=0$ to $M-1$ give
\begin{align*}
\mathbb{E}[L(\M{U}(MT))] - \mathbb{E}[L(\M{U}(0))] \leq CM - \sum_{m=0}^{M-1} \sum_{i,j\in\Set{K}}\theta_{ij}\mathbb{E}[ {U}_{ij}(mT)	]
\end{align*}
which implies that
\begin{align*}
\frac{1}{M}\sum_{m=0}^{M-1}\sum_{i,j\in\Set{K}} \theta_{ij}\mathbb{E}[ {U}_{ij}(mT)	] \leq C +\mathbb{E}[L(\M{U}(0))]/M.
\end{align*}
Then for any $i, j\in\Set{K}$,
\begin{align}
\frac{1}{M}\sum_{m=0}^{M-1} \mathbb{E}[ {U}_{ij}(mT)	]\leq \frac{C +\mathbb{E}[L(\M{U}(0))]/M}{\theta_{ij}}.
\end{align}
Moreover, for any $t_0$, we write $t_0$ as $t_0 = \lfloor t_0/T\rfloor T +\tau$, where $\tau$ is the remainder after diving $t_0$ by $T$, and evidently $\tau\in \{1,2,\dotsc, T-1\}$. Then for $V>TA_\mathrm{max}$
\begin{align}\nonumber
\mathbb{P}[ {U}_{ij}(t_0)>(T+1)V+TA_\mathrm{max}] &\leq \mathbb{P}[ {U}_{ij}( \lfloor t_0/T\rfloor T)>V\\
&\hspace{-15mm}\text{ or }\exists\,\tau_0 \in\{ 1,2,\dotsc,T-1\},{\A}_{ij}( \lfloor t_0/T\rfloor T+\tau_0)> V+A_\mathrm{max} ] \label{eq:t_0_inequality1} \\
&\leq \mathbb{P}[ {U}_{ij}( \lfloor t_0/T\rfloor T)>V]+\frac{TA^2_\mathrm{max}}{V^2}.\label{eq:t_0_inequality2}
\end{align}
The inequality \eqref{eq:t_0_inequality1} is because if $ U(\lfloor t_0/T\rfloor T)\leq V$ and ${\A}_{ij}( \lfloor t_0/T\rfloor T+\tau_1)\leq V+A_\mathrm{max}$ holds for $\tau_1=1,2,\dotsc, T-1$, then 
\begin{align*}
{U}_{ij}(t_0)\leq  {U}_{ij}( \lfloor t_0/T\rfloor T) +\sum_{\tau_1=1}^{\tau}{\A}_{ij}( \lfloor t_0/T\rfloor T+\tau_1)\leq 
(T+1)V+TA_\mathrm{max}.
\end{align*}
The inequality \eqref{eq:t_0_inequality2} is due to the union bound and  \eqref{eq:A_outage_bound}. We then have
\begin{align*}
&\frac{1}{t} \sum_{\tau = 0}^{t-1}\mathbb{P}\big[{\UU}_{ij}(\tau) > (T+1)V+TA_\mathrm{max}\big] \\
&\quad\leq \frac{1}{ \lfloor t/T\rfloor }\sum_{m_0 = 0}^{ \lfloor t/T\rfloor-1}\mathbb{P}\big[{\UU}_{ij}(m_0T)> (T+1)V+TA_\mathrm{max}\big] + \frac{TA^2_\mathrm{max}}{V^2} \\
&\quad\leq \frac{1}{\lfloor t/T\rfloor }\frac{\sum_{m_0 = 0}^{\lfloor t/T\rfloor-1 }\mathbb{E}\big[{\UU}_{ij}(m_0T)\big]}{(T+1)V+TA_\mathrm{max}}+ \frac{TA^2_\mathrm{max}}{V^2} \\
&\quad \leq \frac{1}{\lfloor t/T\rfloor \theta_{ij}}\frac{\lfloor t/T\rfloor C + \mathbb{E}[L(\M{U}(0))]}{(T+1)V+TA_\mathrm{max}}+ \frac{TA^2_\mathrm{max}}{V^2}.
\end{align*}
Taking $\limsup_{t\rightarrow\infty}$, we see that $g_{ij}(V)$ defined in \eqref{eq:g_function} is on the order of $O(1/V)$. Taking limits as $V\rightarrow\infty$, we have $g_{ij}(V)\rightarrow 0$, $i,j\in\Set{K}$.

\section{Proof of Lemma \ref{lemma:T_step_lyapunov_bound}}\label{apd:T_step_Lyapunov_bound}
The $T$-step dynamics of unprocessed entanglement request satisfies
\begin{align*}
\UU_{ij}(t_0+T) & \leq \max\Big\{ \UU_{ij}(t_0) - \Ent_{ij}(t_0) - \sum_{t = t_0}^{t_0+T-1}	 \RF_{ij}(t), 0\Big\} + \sum_{t=t_0}^{t_0+T-1} \A_{ij}(t) \\
&\leq \max\Big\{ \UU_{ij}(t_0) - \sum_{t = t_0}^{t_0+T-1}	 \RF_{ij}(t), 0\Big\} + \sum_{t=t_0}^{t_0+T-1} \A_{ij}(t) 
\end{align*}
where the first inequality is because some entanglement requests during the $T$-slot interval may be processed and the second inequality is because $\Ent_{ij}(t_0)\geq 0$. Squaring both sides and using the fact that $(\max\{x, 0\})^2\leq x^2$ give
\begin{align*}
&\big[\UU_{ij}(t_0+T)\big]^2 - \big[\UU_{ij}(t_0)\big]^2 \\
&\leq \Big[\sum_{t=t_0}^{t_0+T-1}\RF_{ij}(t)\Big]^2 - 2\UU_{ij}(t_0)\sum_{t = t_0}^{t_0+T-1}\RF_{ij}(t)+\Big[ \sum_{t = t_0}^{t_0+T-1}\A_{ij}(t)\Big]^2 \\
&\quad + 2\max\Big\{ \UU_{ij}(t_0) - \sum_{t = t_0}^{t_0+T-1}	 \RF_{ij}(t), 0\Big\} \sum_{t = t_0}^{t_0+T-1}\A_{ij}(t) \\
&\leq  \Big[\sum_{t=t_0}^{t_0+T-1}\RF_{ij}(t)\Big]^2	+ \Big[ \sum_{t = t_0}^{t_0+T-1}\A_{ij}(t)\Big]^2 +2 \UU_{ij}(t_0)\sum_{t=t_0}^{t_0+T-1}\Big[ \A_{ij}(t) - \RF_{ij}(t)\Big].
\end{align*}
Taking conditional expectations on both sides and noting that
\begin{align*}
\mathbb{E}\bigg[ \Big( \sum_{t = t_0}^{t_0+T-1}{\A}_{ij}(t)\Big)^2 \big | \M{U}(t_0)=\V{u} \bigg] \leq T  \sum_{t = t_0}^{t_0+T-1}\mathbb{E}\bigg[ {\A}^2_{ij}(t) \big | \M{U}(t_0)=\V{u} \bigg] \leq T^2 A^2_\mathrm{max}
\end{align*}
we have the desired result.

\section{Proof of Lemma \ref{lemma:lemma_long}}\label{apd:lemma_long}
Since $\V{\Lambda}$ is an interior point of the capacity region, there exists $\delta>0$ such that $\sum_{i\in\Set{K}}\lambda_{ij}< q(1-\delta)p_{j}$, $\forall j$. Set $\epsilon$ to be a positive number such that
\begin{align}\label{eq:select_delta}
\max_{j\in\Set{K}}\frac{2\epsilon}{p_{j}+\epsilon}+q(1-\delta)(1+\delta_\mathrm{max})\frac{p_{j}}{p_{j}+\epsilon} \leq q(1-\delta/2)
\end{align}
where $\delta_{\mathrm{max}} = \max_{i,j}{\epsilon}/{\lambda_{ij}}$. Such an $\epsilon$ must exist because of the mean value theorem.\footnote{Note that the left-hand side is $q(1-\delta)$ and $\infty$ when $\epsilon$ takes the value $0$ and $\infty$, respectively.} We then define the following events:
\begin{align*}
&B_1:\exists i, j,~\bigg|\frac{1}{t+1}\sum_{\tau = 0}^{t}\A_{ij}(\tau)-\lambda_{ij}\bigg|>\epsilon \\
&B_2:\exists j, ~\bigg|\frac{1}{t+1}\sum_{\tau = 0}^{t}\CC_{0j}(\tau)-p_{j}\bigg|>\epsilon \\
& B_3:\exists i,j,~\frac{\sum_{\tau = 0}^{t-1}\RF_{ij}(\tau)}{\sum_{\tau = 0}^{t-1}\F_{ij}(\tau)}\leq q(1-\delta^2/4) \text{ and }\sum_{\tau = 0}^{t-1}\F_{ij}(\tau)\geq \frac{\delta t [(p_{j}-\epsilon) -q(1-\delta)p_{j}(1+\delta_{\mathrm{max}})]}{K(2+\delta)}>0.
\end{align*}
We next bound the set $\overline{S}(t)$.
\begin{lemma}\label{eq:S_bar_B}
The following relationship holds:
\begin{align}\label{eq:S_subset}
\overline{S}(t) \subseteq B_1 \cup B_2 \cup B_3.
\end{align}
\end{lemma}
\begin{IEEEproof}
To show \eqref{eq:S_subset} holds, we only need to show that $\overline{S}(t)\cap \overline{B}_1\cap \overline{B}_2 \subseteq B_3$.

In fact, if $\overline{S}(t)\cap \overline{B}_1\cap \overline{B}_2$ occurs, then the definition of $S(t)$ implies that $\exists j_0$ such that
\begin{align}\label{eq:S_j0}
\sum_{i\in\Set{K}}[\UU_{ij_0}(t)+\A_{ij}(t)]>\Ent_{0j_0}(t)+\CC_{0j_0}(t).
\end{align}
The requirement \eqref{eq:demand_constraint} in on-demand protocols implies that
\begin{align*}
\UU_{ij_0}(t) =\sum_{\tau = 0}^{t-1} \A_{ij_0}(\tau) - \sum_{\tau = 0}^{t-1}\RF_{ij_0}(\tau).
\end{align*}
Together with \eqref{eq:S_j0}, we have
\begin{align}\label{eq:A_greater_C}
\sum_{i\in\Set{K}}\Big[\sum_{\tau = 0}^{t} \A_{ij_0}(\tau) - \sum_{\tau = 0}^{t-1}\RF_{ij_0}(\tau)\Big]> \sum_{\tau = 0}^{t}\CC_{0j_0}(\tau) - \sum_{i\in\Set{K}} \sum_{\tau=0}^{t-1}\F_{ij_0}(\tau).
\end{align}

Then we have
\begin{align}\nonumber
&\sum_{i\in\Set{K}}\sum_{\tau=0}^{t-1}\F_{ij_0}(\tau) - \sum_{i\in\Set{K}}\sum_{\tau=0}^{t-1}\RF_{ij_0}(\tau) \\ \nonumber
&> (p_{j_0}-\epsilon)(t+1) - \sum_{i\in\Set{K}}(\lambda_{ij_0}+\epsilon)(t+1)\\ \nonumber
&>(p_{j_0}-\epsilon)(t+1) -(1+\delta_{\mathrm{max}}) \sum_{i\in\Set{K}}\lambda_{ij_0}(t+1) \\ \label{eq:later_use}
&>(p_{j_0}-\epsilon)(t+1) -q(1-\delta)p_{j_0}(1+\delta_{\mathrm{max}})(t+1)
\\
&> \bigg[\frac{p_{j_0}-\epsilon}{p_{j_0}+\epsilon}-q(1-\delta)(1+\delta_\mathrm{max})\frac{p_{j_0}}{p_{j_0}+\epsilon}\bigg]\sum_{\tau = 0}^{t}\CC_{0j_0}(\tau)  \nonumber\\
&\geq  \bigg[\frac{p_{j_0}-\epsilon}{p_{j_0}+\epsilon}-q(1-\delta)(1+\delta_\mathrm{max})\frac{p_{j_0}}{p_{j_0}+\epsilon}\bigg]\sum_{i\in\Set{K}}\sum_{\tau = 0}^{t-1}\F_{ij_0}(\tau) \label{eq:mu_is_small}
\end{align}
where the first inequality is obtained by reordering \eqref{eq:A_greater_C} and using the fact that $\overline{B}_1\cap\overline{B}_2$ occurs, the third inequality is because $\sum_{i\in\Set{K}}\lambda_{ij_0}< q(1-\delta)p_{j_0}$, and the fourth inequality is because $\overline{B}_2$ occurs and
\begin{align}\nonumber
\frac{p_{j_0}-\epsilon}{p_{j_0}+\epsilon}-q(1-\delta)(1+\delta_\mathrm{max})\frac{p_{j_0}}{p_{j_0}+\epsilon} &= 1 - \bigg[	 \frac{2\epsilon}{p_{j_0}+\epsilon}+q(1-\delta)(1+\delta_\mathrm{max})\frac{p_{j_0}}{p_{j_0}+\epsilon} \bigg] \\ \label{eq:positive_for_bound}
&\geq 1-q(1-\delta/2)\\
&> 1-q>0 \nonumber
\end{align}
in which \eqref{eq:positive_for_bound} is due to \eqref{eq:select_delta}. We next show that if \eqref{eq:later_use} and \eqref{eq:mu_is_small} hold, then $B_3$ occurs, showing that $\overline{S}(t)\cap \overline{B}_1\cap \overline{B}_2 \subseteq B_3$ and the proof is finished.

In fact, \eqref{eq:select_delta} and \eqref{eq:mu_is_small} and imply
\begin{align*}
q(1-\delta/2)\sum_{i\in\Set{K}}\sum_{\tau = 0}^{t-1}\F_{ij_0}(\tau) \geq \sum_{i\in\Set{K}} \sum_{\tau = 0}^{t-1}\RF_{ij_0}(\tau).
\end{align*}
Using the following lemma with 
\begin{align*}
a_i &= \sum_{\tau = 0}^{t-1}\F_{ij_0}(\tau)\\
b_i&=\sum_{\tau = 0}^{t-1}\RF_{ij_0}(\tau)\\
c &=q(1-\delta/2) \\
d&=(p_{j_0}-\epsilon)t -q(1-\delta)p_{j_0}(1+\delta_{\mathrm{max}})t \\
\delta_0& = \delta/2
\end{align*}
and we have the desired result that $B_3$ occurs.

\begin{lemma}\label{lemma:one_from_K}
For $a_i,b_i\in \mathbb{R}_{\geq 0}$, $i=1,2,\dots, K$ and $c, d>0$, if 
\begin{align}
c\sum_{i=1}^K a_i &\geq \sum_{i=1}^K b_i \\
a_i &\geq b_i, ~~i=1,2,\dots, K \\
\sum_{i=1}^K a_i &\geq d \label{eq:sum_a_i}
\end{align}
then for any $\delta_0>0$, there exists $i_0$ such that
\begin{align*}
(1+\delta_0)c a_{i_0}&\geq b_{i_0} \\
a_{i_0}&\geq \frac{ d\delta_0}{K (1+\delta_0)}.
\end{align*}
\end{lemma}
\begin{IEEEproof}
Suppose such $i_0$ does not exist. Then for any $i$, $(1+\delta_0 )ca_i<b_i$ or $a_i< { d\delta_0}/({K (1+\delta_0)})$. Then for any $i$,
\begin{align*}
(1+\delta_0 )ca_i < b_i + \frac{  cd\delta_0  }{ K }.
\end{align*}
Summing over $i$, we have
\begin{align*}
\sum_{i=1}^K(1+\delta_0 )ca_i & <\Big(\sum_{i=1}^K b_i\Big) +cd\delta_0\\
&\leq \Big(\sum_{i=1}^K ca_i \Big)+cd \delta_0\end{align*}
implying that $\sum_{i=1}^K a_i <d.$ This contradicts the condition \eqref{eq:sum_a_i} and finishes the proof.
\end{IEEEproof}

 Together with \eqref{eq:select_delta}, this implies that $B_3$ occurs, and \eqref{eq:S_subset} follows.
\end{IEEEproof}

We next prove \eqref{eq:cpnditional_exp_A}. Note that
\begin{align}\nonumber
\mathbb{E}\Big[ \sum_{\tau = 0}^{t}\A_{\tilde{i}\tilde{j}}(\tau)|\overline{S}(t)\Big]\mathbb{P}\big[\overline{S}(t)\big] & = \sum_{a = 0}^\infty a \mathbb{P}\Big[\sum_{\tau = 0}^{t}\A_{\tilde{i}\tilde{j}}(\tau) = a, \overline{S}(t)\Big] \\ \nonumber
&\leq  \sum_{a = 0}^\infty a \mathbb{P}\Big[\sum_{\tau = 0}^{t}\A_{\tilde{i}\tilde{j}}(\tau) = a, B_1\cup B_2 \cup (\overline{S}(t)\cap B_3)\Big] \\ \nonumber
&\leq \sum_{a = 0}^\infty a \mathbb{P}\Big[\sum_{\tau = 0}^{t}\A_{\tilde{i}\tilde{j}}(\tau) = a, B_1\Big] + \sum_{a = 0}^\infty a \mathbb{P}\Big[\sum_{\tau = 0}^{t}\A_{\tilde{i}\tilde{j}}(\tau) = a, B_2\Big] \\
&\quad + \sum_{a = 0}^\infty a \mathbb{P}\Big[\sum_{\tau = 0}^{t}\A_{\tilde{i}\tilde{j}}(\tau) = a,\overline{B}_1\cap \overline{B}_2\cap \overline{S}(t)\cap B_3\Big] \label{eq:three_terms}
\end{align}
where the first inequality is because of \eqref{eq:S_subset} and the second inequality is because for any events $A$, $B$, and $C$, $\mathbb{P}[A\cup B\cup C]\leq \mathbb{P}[A]+\mathbb{P}[B]+\mathbb{P}[\overline{A}\cap\overline{B}\cap {C}]$.

We then provide upper bounds for each term in \eqref{eq:three_terms}. Note that
\begin{align}\label{eq:first_bound}
\sum_{a = 0}^\infty a \mathbb{P}\Big[\sum_{\tau = 0}^{t}\A_{{\tilde{i}\tilde{j}}}(\tau) = a, B_1\Big] &= \mathbb{E}\Big[\sum_{\tau = 0}^{t}{\A}_{{\tilde{i}\tilde{j}}}(\tau) \big | B_1\Big] \mathbb{P}[B_1] \leq c_1(\epsilon)
\end{align}
where the inequality is because of the assumption \eqref{eq:assumption_Aij}. For the second term in \eqref{eq:three_terms}, by applying the
Chernoff-Hoeffding theorem, we have
\begin{align}\nonumber
 &\sum_{a = 0}^\infty a \mathbb{P}\Big[\sum_{\tau = 0}^{t}\A_{{\tilde{i}\tilde{j}}}(\tau) = a, B_2\Big] \\ \nonumber
 &= 
 \mathbb{E}\Big[\sum_{\tau = 0}^{t}\A_{{\tilde{i}\tilde{j}}}(\tau)\Big] \mathbb{P}[B_2] 
 \\ \label{eq:second_bound}
 &\leq  \lambda_{{\tilde{i}\tilde{j}}}t \sum_{j\in\Set{K}}\big[
 \exp\{-D(p_{j}+\epsilon||p_{j})t\}+\exp\{-D(p_{j}-\epsilon||p_{j})t\}\big]
\end{align}
where
\begin{align}\label{eq:Divergence}
D(x||y) = x\ln\frac{x}{y}+(1-x)\ln\Big(\frac{1-x}{1-y}\Big).
\end{align}
Now consider the last term in \eqref{eq:three_terms}. 
\begin{align*}
&\mathbb{P}[\overline{B}_1\cap \overline{B}_2\cap \overline{S}(t)\cap B_3] \leq \mathbb{P}[ B_3 ] \\
&\leq \frac{K^2}{2}\exp\Big\{-   \frac{D(q-q\delta^2/4||q)\delta t }{K(2+\delta)}  \min_{j\in{\Set{K}}}[(p_{j}-\epsilon) -q(1-\delta)p_{j}(1+\delta_{\mathrm{max}})] \Big\}
\end{align*}
where the second inequality is because of the Chernoff-Hoeffding theorem and the coefficient $K^2/2$ is because of the union bound. We then have
\begin{align}\nonumber
& \sum_{a = 0}^\infty a \mathbb{P}\Big[\sum_{\tau = 0}^{t-1}\A_{{\tilde{i}\tilde{j}}}(\tau) = a,\overline{B}_1\cap \overline{B}_2\cap \overline{S}(t)\cap B_3\Big] \\ \nonumber
 &\leq (\lambda_{{\tilde{i}\tilde{j}}}+\epsilon)t\sum_{ (\lambda_{{\tilde{i}\tilde{j}}}-\epsilon)t\leq a \leq  (\lambda_{{\tilde{i}\tilde{j}}}+\epsilon)t}  \mathbb{P}\Big[\sum_{\tau = 0}^{t-1}\A_{{\tilde{i}\tilde{j}}}(\tau) = a,\overline{B}_1\cap \overline{B}_2\cap \overline{S}(t)\cap B_3\Big] \\ \nonumber
 &= (\lambda_{{\tilde{i}\tilde{j}}}+\epsilon)t\mathbb{P}\Big[\overline{B}_1\cap \overline{B}_2\cap \overline{S}(t)\cap B_3\Big]\\
 &\leq  \frac{K^2(\lambda_{{\tilde{i}\tilde{j}}}+\epsilon)t}{2}\exp\Big\{-   \frac{D(q-q\delta^2/4||q)\delta t }{K(2+\delta)}  \min_{j\in{\Set{K}}}[(p_{j}-\epsilon) -q(1-\delta)p_{j}(1+\delta_{\mathrm{max}})] \Big\}
 \label{eq:third_bound}
\end{align}
where the first inequality is because $\overline{B}_1$ occurs.

We now find the upper bounds \eqref{eq:first_bound}, \eqref{eq:second_bound}, and \eqref{eq:third_bound} for the terms in \eqref{eq:three_terms}. Note that \eqref{eq:second_bound} and \eqref{eq:third_bound} converge to 0 when $t$ goes to infinity, and therefore can be further upper-bounded by a constant irrelevant of $t$. This completes the proof.

\section{Proof of Lemma \ref{lemma:lemma_long_virtual}}\label{apd:lemma_long_virtual}
Since $\V{\Lambda}$ is an interior point of the capacity region, there exists $\delta>0$ such that $\sum_{i\in\Set{K}}\lambda_{ij}< q(1-\delta)p_{j}$, $\forall j$. Consider $t$ such that
\begin{align}\label{eq:t_select}
t\geq \max_{j\in \Set{K}} {4K}/(p_j \delta).
\end{align}
Set $\tilde{\epsilon}\in (0,\min_{j\in\Set{K}}p_j)$ to be a positive number such that
\begin{align}\label{eq:select_delta_virtual}
\min_{j\in\Set{K}}-\frac{Kq}{(t+1)(p_{j}-\tilde{\epsilon})}+q-q(1-\delta)(1+\tilde{\delta}_\mathrm{max})\frac{p_{j}}{p_{j}-\tilde{\epsilon}} \geq q \delta/2
\end{align}
where $\tilde{\delta}_{\mathrm{max}} = \max_{i,j}{\tilde{\epsilon}}/{\lambda_{ij}}$. Such an $\tilde{\epsilon}$ must exist because when $\epsilon \rightarrow 0^+$, the left side of \eqref{eq:select_delta_virtual} is greater than $q\delta/2$ due to the choice of $t$ in \eqref{eq:t_select}.

We then define the following events:
\begin{align*}
&C_1:\exists i, j,~ \frac{1}{t+1}\sum_{\tau = 0}^{t}\WA_{ij}(\tau) >\lambda_{ij} +\tilde{\epsilon} \\
&C_2:\exists j, ~\bigg|\frac{1}{t+1}\sum_{\tau = 0}^{t}\CC_{0j}(\tau)-p_{j}\bigg|>\tilde{\epsilon} \\
& C_3:\exists i,j,~\frac{\sum_{\tau = 0}^{t-1}\RF_{ij}(\tau)}{\sum_{\tau = 0}^{t-1}\F_{ij}(\tau)}\leq q(1-\delta^2/4) \text{ and }\sum_{\tau = 0}^{t-1}\F_{ij}(\tau)\geq \frac{q\delta^2  (p_{j}-\tilde{\epsilon})(t+1)}{2K(2+\delta)}.
\end{align*}
We next bound the set $\overline{S}_0(t)$.
\begin{lemma}\label{eq:S_bar_B_virtual}
The following relationship holds:
\begin{align}\label{eq:S_subset_virtual}
\overline{S}_0(t) \subseteq C_1 \cup C_2 \cup C_3.
\end{align}
\end{lemma}
\begin{IEEEproof}
To show \eqref{eq:S_subset_virtual} holds, we only need to show that $\overline{S}_0(t)\cap \overline{C}_1\cap \overline{C}_2 \subseteq C_3$.

In fact, if $\overline{S_0}(t)\cap \overline{C}_1\cap \overline{C}_2$ occurs, then the definition of $S_0(t)$ implies that $\exists j_0$ such that
\begin{align}\label{eq:S_j0_virtual}
\sum_{i\in\Set{K}}\Big\lceil	\big[{\WU}_{ij_0}(t)+\WA_{ij_0}(t)\big]/q\Big\rceil>\Ent_{0j_0}(t)+\CC_{0j_0}(t).
\end{align}
The definition of $\WU_{ij}(t)$ in on-demand protocols with virtual requests implies that
\begin{align*}
\WU_{ij_0}(t) =\sum_{\tau = 0}^{t-1} \WA_{ij_0}(\tau) - \sum_{\tau = 0}^{t-1}\RF_{ij_0}(\tau).
\end{align*}
Together with \eqref{eq:S_j0_virtual}, we have
\begin{align}\label{eq:A_greater_C_virtual}
\sum_{i\in\Set{K}}\Big\{1+\Big[\sum_{\tau = 0}^{t} \WA_{ij_0}(\tau) - \sum_{\tau = 0}^{t-1}\RF_{ij_0}(\tau)\Big]/q\Big\}> \sum_{\tau = 0}^{t}\CC_{0j_0}(\tau) - \sum_{i\in\Set{K}} \sum_{\tau=0}^{t-1}\F_{ij_0}(\tau).
\end{align}

Then we have
\begin{align}\nonumber
q\sum_{i\in\Set{K}} \sum_{\tau=0}^{t-1}\F_{ij_0}(\tau) &> \sum_{i\in\Set{K}}\sum_{\tau = 0}^{t-1}\RF_{ij_0}(\tau)-Kq+q\sum_{\tau = 0}^{t}\CC_{0j_0}(\tau)-\sum_{i\in\Set{K}}\sum_{\tau = 0}^{t} \WA_{ij_0}(\tau) \\ \nonumber
&\geq \sum_{i\in\Set{K}}\sum_{\tau = 0}^{t-1}\RF_{ij_0}(\tau)-Kq+q\sum_{\tau = 0}^{t}\CC_{0j_0}(\tau)-(1+\tilde{\delta}_{\mathrm{max}})(t+1) \sum_{i\in\Set{K}}\lambda_{ij_0} \\ \nonumber
&\geq \sum_{i\in\Set{K}}\sum_{\tau = 0}^{t-1}\RF_{ij_0}(\tau)-Kq+q\sum_{\tau = 0}^{t}\CC_{0j_0}(\tau)-q(1+\tilde{\delta}_{\mathrm{max}})(t+1)(1-\delta)p_{j_0} \\ \nonumber
&\geq \sum_{i\in\Set{K}}\sum_{\tau = 0}^{t-1}\RF_{ij_0}(\tau)-Kq+q\sum_{\tau = 0}^{t}\CC_{0j_0}(\tau)-\frac{q(1+\tilde{\delta}_{\mathrm{max}}) (1-\delta)p_{j_0}}{p_{j_0}-\tilde{\epsilon}}\sum_{\tau = 0}^{t}\CC_{0j_0}(\tau)
\\  \label{eq:total_F_virtual}
&\geq \sum_{i\in\Set{K}}\sum_{\tau = 0}^{t-1}\RF_{ij_0}(\tau) +q\delta/2 \sum_{\tau = 0}^{t}\CC_{0j_0}(\tau)	\\
&\geq \sum_{i\in\Set{K}}\sum_{\tau = 0}^{t-1}\RF_{ij_0}(\tau) +q\delta/2\sum_{i\in\Set{K}} \sum_{\tau=0}^{t-1}\F_{ij_0}(\tau) \label{eq:mu_is_small_virtual}
\end{align}
where the first inequality by reordering \eqref{eq:A_greater_C_virtual}, the second inequality is because $\overline{C}_1$ occurs, the third inequality is because $\sum_{i\in\Set{K}}\lambda_{ij_0}< q(1-\delta)p_{j_0}$, the fourth inequality is because $\overline{C}_2$ occurs, and the fifth inequality is because of \eqref{eq:select_delta_virtual}. Equations \eqref{eq:total_F_virtual} and \eqref{eq:mu_is_small_virtual} imply
\begin{align}\label{eq:F_mu_ratio_outlier}
q(1-\delta/2)\sum_{i\in\Set{K}}\sum_{\tau = 0}^{t-1}\F_{ij_0}(\tau) &\geq \sum_{i\in\Set{K}} \sum_{\tau = 0}^{t-1}\RF_{ij_0}(\tau)\\ \label{eq:total_F_abs_virtual}
\sum_{i\in\Set{K}}\sum_{\tau = 0}^{t-1}\F_{ij_0}(\tau) &\geq q(p_{j_0}-\epsilon)(t+1)\delta/2.
\end{align}

If \eqref{eq:F_mu_ratio_outlier} and \eqref{eq:total_F_abs_virtual} hold, then $C_3$ occurs because of  Lemma \ref{lemma:one_from_K} with
\begin{align*}
a_i &= \sum_{\tau = 0}^{t-1}\F_{ij_0}(\tau)\\
b_i&=\sum_{\tau = 0}^{t-1}\RF_{ij_0}(\tau)\\
c &=q(1-\delta/2) \\
d&= q(p_{j_0}-\epsilon)(t+1)\delta/2 \\
\delta_0& = \delta/2.
\end{align*}

 This shows that $\overline{S}(t)\cap \overline{C}_1\cap \overline{C}_2 \subseteq C_3$ and the proof of Lemma \ref{eq:S_bar_B_virtual} is finished.
 \end{IEEEproof}

We next prove \eqref{eq:conditional_exp_A0_virtual}. Note that
\begin{align}\nonumber
\mathbb{E}\Big[ \sum_{\tau = 0}^{t}\WA_{\tilde{i}\tilde{j}}(\tau)|\overline{S_0}(t)\Big]\mathbb{P}\big[\overline{S_0}(t)\big] & = \sum_{a = 0}^\infty a \mathbb{P}\Big[\sum_{\tau = 0}^{t}\WA_{\tilde{i}\tilde{j}}(\tau) = a, \overline{S_0}(t)\Big] \\ \nonumber
&\leq  \sum_{a = 0}^\infty a \mathbb{P}\Big[\sum_{\tau = 0}^{t}\WA_{\tilde{i}\tilde{j}}(\tau) = a, C_1\cup C_2 \cup (\overline{S_0}(t)\cap C_3)\Big] \\ \nonumber
&\leq \sum_{a = 0}^\infty a \mathbb{P}\Big[\sum_{\tau = 0}^{t}\WA_{\tilde{i}\tilde{j}}(\tau) = a, C_1\Big] + \sum_{a = 0}^\infty a \mathbb{P}\Big[\sum_{\tau = 0}^{t}\WA_{\tilde{i}\tilde{j}}(\tau) = a, C_2\Big] \\
&\quad + \sum_{a = 0}^\infty a \mathbb{P}\Big[\sum_{\tau = 0}^{t}\WA_{\tilde{i}\tilde{j}}(\tau) = a,\overline{C}_1\cap \overline{C}_2\cap \overline{S}(t)\cap C_3\Big] \label{eq:three_terms_virtual}
\end{align}
where the first inequality is because of \eqref{eq:S_subset_virtual} and the second inequality is because for any events $A$, $B$, and $C$, $\mathbb{P}[A\cup B\cup C]\leq \mathbb{P}[A]+\mathbb{P}[B]+\mathbb{P}[\overline{A}\cap\overline{B}\cap {C}]$.

We then show each term in \eqref{eq:three_terms_virtual} converges to 0 as $t$ goes to infinity. Note that
\begin{align}\nonumber
&\lim_{t\rightarrow \infty}\sum_{a = 0}^\infty a \mathbb{P}\Big[\sum_{\tau = 0}^{t}\WA_{{\tilde{i}\tilde{j}}}(\tau)= a, C_1\Big] \\
&=\lim_{t\rightarrow \infty} \mathbb{E}\Big[\lceil t^\alpha\rceil+\sum_{\tau = 0}^{t}{\A}_{{\tilde{i}\tilde{j}}}(\tau) \big | C_1\Big] \mathbb{P}[C_1] \nonumber \\
&=\Big(\lim_{t\rightarrow \infty} \lceil t^\alpha\rceil  \mathbb{P}[C_1] \Big)+\Big(\lim_{t\rightarrow \infty} \mathbb{E}\Big[\sum_{\tau = 0}^{t}{\A}_{{\tilde{i}\tilde{j}}}(\tau) \big | C_1\Big] \mathbb{P}[C_1] \Big)  \label{eq:first_bound_virtual}\\
&=0\nonumber
\end{align}
where in \eqref{eq:first_bound_virtual}, the first term converges to 0 because $\mathbb{P}[C_1]\sim o(1/t)$ due to Assumption 2, and the second term converges to 0 due to Assumption 2 and the fact that $\lceil t^\alpha\rceil/(t+1)\xrightarrow{t\rightarrow \infty} 0$.

For the second term in \eqref{eq:three_terms_virtual}, by applying the
Chernoff-Hoeffding theorem, we have
\begin{align}\nonumber
 &\sum_{a = 0}^\infty a \mathbb{P}\Big[\sum_{\tau = 0}^{t}\WA_{{\tilde{i}\tilde{j}}}(\tau) = a, C_2\Big] \\ \nonumber
 &= 
 \mathbb{E}\Big[\sum_{\tau = 0}^{t}\WA_{{\tilde{i}\tilde{j}}}(\tau)\Big] \mathbb{P}[C_2] 
 \\ \label{eq:second_bound_virtual}
 &\leq  \lambda_{{\tilde{i}\tilde{j}}}t \sum_{j\in\Set{K}}\big[
 \exp\{-D(p_{j}+\tilde{\epsilon}||p_{j})t\}+\exp\{-D(p_{j}-\tilde{\epsilon}||p_{j})t\}\big].
\end{align}


Now consider the last term in \eqref{eq:three_terms_virtual}. 
\begin{align*}
&\mathbb{P}[\overline{C}_1\cap \overline{C}_2\cap \overline{S_0}(t)\cap C_3] \leq \mathbb{P}[ C_3 ] \\
&\leq \frac{K^2}{2}\exp\Big\{-   \frac{q\delta^2(\min_{j\in\Set{K}} p_j-\tilde{\epsilon})(t+1)D(q-q\delta^2/4||q) }{2K(2+\delta)}  \Big\}
\end{align*}
where the second inequality is because of the Chernoff-Hoeffding theorem and the coefficient $K^2/2$ is because of the union bound. We then have
\begin{align}\nonumber
& \sum_{a = 0}^\infty a \mathbb{P}\Big[\sum_{\tau = 0}^{t-1}\WA_{{\tilde{i}\tilde{j}}}(\tau) = a,\overline{C}_1\cap \overline{C}_2\cap \overline{S_0}(t)\cap C_3\Big] \\ \nonumber
 &\leq (\lambda_{{\tilde{i}\tilde{j}}}+\tilde{\epsilon})t\sum_{  a \leq  (\lambda_{{\tilde{i}\tilde{j}}}+\tilde{\epsilon})t}  \mathbb{P}\Big[\sum_{\tau = 0}^{t-1}\WA_{{\tilde{i}\tilde{j}}}(\tau) = a,\overline{C}_1\cap \overline{C}_2\cap \overline{S_0}(t)\cap C_3\Big] \\ \nonumber
 &= (\lambda_{{\tilde{i}\tilde{j}}}+\tilde{\epsilon})t\, \mathbb{P}\Big[\overline{C}_1\cap \overline{C}_2\cap \overline{S_0}(t)\cap C_3\Big]\\
 &\leq  \frac{K^2(\lambda_{{\tilde{i}\tilde{j}}}+\tilde{\epsilon})t}{2}\exp\Big\{-   \frac{q\delta^2(\min_{j\in\Set{K}} p_j-\tilde{\epsilon})(t+1)D(q-q\delta^2/4||q) }{2K(2+\delta)}  \Big\}
 \label{eq:third_bound_virtual}
\end{align}
where the first inequality is because $\overline{C}_1$ occurs.

Combining \eqref{eq:first_bound_virtual}, \eqref{eq:second_bound_virtual}, and \eqref{eq:third_bound_virtual}, we see that the right hand side of \eqref{eq:three_terms_virtual} converges to 0 when $t$ goes to infinity. This completes the proof of Lemma \ref{lemma:lemma_long_virtual}.

\section{Proof of Convergence in \eqref{eq:new_three_term}}\label{apd:new_three_term}
The first term in \eqref{eq:new_three_term}  converges to 0 because
\begin{align*}
 \mathbb{E}\big[\max\{\WU_{ij}(t+1)-\lceil (t+1)^\alpha\rceil, 0\}| C_1\big] \mathbb{P}\big[C_1\big] &\leq \mathbb{E}\big[ \WU_{ij}(t+1) | C_1\big] \mathbb{P}\big[C_1\big]\\
 &\leq \mathbb{E}\Big[ \sum_{\tau = 0}^{t}\WA_{ij}(\tau)|C_1\Big]\mathbb{P}\big[C_1\big]
 \end{align*}
 which converges to 0 as shown in \eqref{eq:first_bound_virtual}.
 
 We now consider the second term in \eqref{eq:new_three_term}. The second term is a conditional expectation when the event  ${S_0}(t)\cap \overline{C}_1$ occurs. Note that $\WU_{ij}(t+1)= \WU_{ij}(t)+\WA_{ij}(t)-\RF_{ij}(t)$.  Let 
 \begin{align*}
 D_1=\big\{\WU_{ij}(t)+\WA_{ij}(t)> \lceil (t+1)^\alpha \rceil\big\}.
 \end{align*}  
 If $\overline{D}_1$ occurs, then $
 \max\{\WU_{ij}(t+1)-\lceil (t+1)^\alpha\rceil, 0\} =0$.  We next consider that $D_1$ occurs. We have $F_{ij}(t)=\Big\lceil \big[\WU_{ij}(t)+\WA_{ij}(t)\big]/q\Big\rceil$ because $S_0$ occurs. Let $\beta \in (0, \alpha-0.5)$. Let 
  \begin{align*}
 D_2=\big\{	\WU_{ij}(t)+\WA_{ij}(t)-\RF_{ij}(t) &\geq \big[\big(\WU_{ij}(t)+\WA_{ij}(t)\big)/q\big]^{0.5+\beta}	\big\}.
 \end{align*} 
 
Note that $\mathbb{P} [D_2 \,|\,D_1 ]$ can be bounded by
 \begin{align*}
\mathbb{P} [D_2 \,|\,D_1 ]
&\leq 2\exp\Big\{-2\big[(\WU_{ij}(t)+\WA_{ij}(t))/q\big]^{2\beta}\Big\}\\
&\leq 2\exp{-2(t+1)^{\alpha\beta} }
 \end{align*}
 where the first inequality is because of Chernoff's bound and the last inequality is because ${D}_1$ occurs. If $\overline{D}_2$ occurs, then
 \begin{align*}
 \max\{\WU_{ij}(t+1)-\lceil (t+1)^\alpha\rceil, 0\}&\leq \max\Big\{ \big[  \big(\WU_{ij}(t)+\WA_{ij}(t)\big)/q\big]^{0.5+\beta}-\lceil (t+1)^\alpha\rceil, 0\Big\}\\
 &\leq  \max \Big\{\big[(\lambda_{ij}+\tilde{\epsilon})t/q\big]^{0.5+\beta}-\lceil (t+1)^\alpha \rceil, 0\Big\}
 \end{align*}
and consequently,
 \begin{align*}
&  \mathbb{E}\big[\max\{\WU_{ij}(t+1)-\lceil (t+1)^\alpha\rceil, 0\}|  {S_0}(t)\cap \overline{C}_1\big] \mathbb{P}\big[ {S_0}(t)\cap \overline{C}_1\big]\\
& =\mathbb{E}\big[\max\{\WU_{ij}(t+1)-\lceil (t+1)^\alpha\rceil, 0\}|  {S_0}(t)\cap \overline{C}_1\cap D_1\big] \mathbb{P}\big[ {S_0}(t)\cap \overline{C}_1\cap D_1\big]\\
&=\mathbb{E}\big[\max\{\WU_{ij}(t+1)-\lceil (t+1)^\alpha\rceil, 0\}|  {S_0}(t)\cap \overline{C}_1\cap D_1\cap D_2\big] \mathbb{P}\big[ {S_0}(t)\cap \overline{C}_1\cap D_1\cap D_2\big]\\
&\quad+\mathbb{E}\big[\max\{\WU_{ij}(t+1)-\lceil (t+1)^\alpha\rceil, 0\}|  {S_0}(t)\cap \overline{C}_1\cap D_1\cap \overline{D}_2\big] \mathbb{P}\big[ {S_0}(t)\cap \overline{C}_1\cap D_1\cap \overline{D}_2\big]\\
&\leq 2(\lambda_{ij}+\tilde{\epsilon})t \exp{-2(t+1)^{\alpha\beta} }+\max \Big\{\big[(\lambda_{ij}+\tilde{\epsilon})t/q\big]^{0.5+\beta}-\lceil (t+1)^\alpha \rceil, 0\Big\}
 \end{align*}
 which converges to 0 as $t$ goes to infinity because $\beta<\alpha-0.5$. This shows that the last term in \eqref{eq:new_three_term} converges to 0.

\end{document}